\documentclass[journal,11pt,onecolumn]{IEEEtran}

\usepackage{amsmath}
\usepackage{amsfonts}
\usepackage{amssymb}
\usepackage{epsfig}
\usepackage{graphicx}
\usepackage{cite}
\usepackage{verbatim}

\newtheorem{theorem}{Theorem}
\newtheorem{lemma}{Lemma}

\newtheorem{proposition}{Proposition}

\title{D-MG Tradeoff of DF and AF Relaying Protocols over Asynchronous PAM Cooperative Networks}

\author{\authorblockN{Mehdi Torbatian and Mohamed Oussama Damen}\\
\authorblockA{Department of Electrical and Computer Engineering\\
University of Waterloo\\
Waterloo, Ontario, Canada\\
Email: \{mtorbatian, mdamen\}@uwaterloo.ca}}
\date{}
\begin{document}

\maketitle

\begin{abstract}
The diversity multiplexing tradeoff of a general two-hop \textit{asynchronous} cooperative network is examined for various relaying protocols such as non-orthogonal selection decode-and-forward (NSDF), orthogonal selection decode-and-forward (OSDF), non-orthogonal amplify-and-forward (NAF), and orthogonal amplify-and-forward (OAF). The transmitter nodes are assumed to send pulse amplitude modulation (PAM) signals asynchronously, in which information symbols are linearly modulated by a shaping waveform to be sent to the destination. We consider two different cases with respect to the length of the shaping waveforms in the time domain. In the theoretical case where the shaping waveforms with infinite time support are used, it is shown that asynchronism does not affect the DMT performance of the system and the same DMT as that of the corresponding synchronous network is obtained for all the aforementioned protocols. In the practical case where finite length shaping waveforms are used, it is shown that better diversity gains can be achieved at the expense of bandwidth expansion. In the decode-and-forward (DF) type protocols, the asynchronous network provides better diversity gains than those of the corresponding synchronous network throughout the range of the multiplexing gain. In the amplify-and-forward (AF) type protocols, the asynchronous network provides the same DMT as that of the corresponding synchronous counterpart under the OAF protocol; however, a better diversity gain is achieved under the NAF protocol throughout the range of the multiplexing gain. In particular, in the single relay asynchronous network, the NAF protocol provides the same DMT as that of the $2 \times 1$ multiple-input single-output (MISO) channel.
\end{abstract}


\section{Introduction}\label{INTRODUCTION}
Cooperative diversity was first proposed as a \textit{synchronous} technique \cite{AazhangI,AazhangII} to provide spatial diversity with the help of surrounding terminals. However, because the relays are at different locations (i.e., different propagation delays) and they have their own local oscillators with no common timing reference, it is an \emph{asynchronous} technique in nature. Although the relays may be synchronized by an infrastructure service provider, this causes a large amount of overhead on the overall throughput of the network.

While previously proposed space-time codes are adapted to use in synchronous cooperative scenarios \cite{Jing-Hassibi,Jing-Jafarkhani}, they cannot realize the capabilities of this technique when they apply to practical asynchronous cases. In contrast, many distributed space-time schemes have been proposed to provide cooperative diversity gains in the presence of the asynchronizm among the relays \cite{Torbatian-Damen,Shang-Xia2,Damen-Hommons}. A common assumption in all of them is that the asynchronous delays are integer factors of the symbol interval and fractional delays (i.e., the non-integer part of the delay) are absorbed in multi path. Such an assumption is reasonable when the fractional delays are very small compared to the length of a symbol interval. Another approach consists in using orthogonal frequency division multiplexing (OFDM) to combat synchronization errors \cite{Mie-Hua,Rajan-Rajan}. In contrast to the previous schemes, OFDM allows the synchronization error to be any factor of the symbol interval.

Contrary to intuition, some exceptions have been reported wherein the asynchronism has helped to improve the system performance \cite{Barman-Dabeer,Verdu,Winters,
Rusek-Anderson,Bossert-Huebner,Wang-Chang-Yang}. For example in \cite{Barman-Dabeer}, it is shown that asynchronous pulse amplitude modulation (PAM) can exploit the total existing degrees of freedom (DOF) of a multiple-input multiple-output (MIMO) system which communicates over a spectral mask with infinite support, while the synchronous PAM exploits only finite number of the DOF of this channel.

In \cite{Shuangqing_Wei}, the effect of the asynchronism on the diversity multiplexing tradeoff (DMT) \cite{Zheng-Tse} of an orthogonal decode-and-forward cooperative network consisting of two parallel relays is examined, in which the transmitting nodes use shaping waveforms spanned over two symbol intervals. The author shows that for large length codewords, the same DMT performance as that of the corresponding synchronous network is achieved. Moreover, when both relays can fully decode the source message, the equivalent channel from the relays to the destination at high values of signal to noise ratio (SNR) behaves similar to a parallel channel with two independent links. The outage probability and the DMT of an asynchronous parallel relay network containing two relays without the direct source-destination link are considered in \cite{Nahas-Saadani-Hachem}. It is shown that the same DMT performance as that of the corresponding synchronous network is achieved. In \cite{Krishnakumar-Naveen-Sreeram-Kumar}, under the assumption of having integer delays, two different models of asynchronism in a cooperative relay network with at least two relays are considered. For each model, a variant of the slotted amplify-and-forward (SAF) relaying protocol \cite{Yang-Belfiore} is proposed which asymptotically achieves the transmit diversity bound in the absence of a direct source-destination link. In the presence of this link, it is shown in \cite{Krishnakumar-Naveen-Kumar} that the SAF protocol is asymptotically optimal under both models of asynchronism. It is worth nothing that in the SAF protocol, the relays are assumed to be isolated from each other which is in fact often unrealistic. For a \textit{synchronous} two-hop cooperative relay network with arbitrary number of relays, the DMT performance is calculated in \cite{Elia-Vinodh-Anand-Kumar} for various relaying protocols such as the orthogonal and non-orthogonal selection decode-and-forward (OSDF and NSDF) and the orthogonal amplify-and-forward (OAF). In each case, a DMT optimal code is constructed using cyclic division algebra space-time codes \cite{Tavildar-Viswanath,Sethuraman-Rajan-Shashidhar,Elia-Sethuraman-Kumar}.
It is shown that by allowing the source and the relays to transmit over proper asymmetric portions of a cooperative frame, a larger diversity gain may be achieved at each multiplexing gain.

In this work, we analyze the DMT performance of a general two-hop \textit{asynchronous} cooperative network containing one source node, one destination node, and $M$ parallel relay nodes for various relaying protocols such as the OSDF, NSDF, OAF, and non-orthogonal amplify-and-forward (NAF). Similar to \cite{Elia-Vinodh-Anand-Kumar}, we let the source and the relays to transmit over asymmetric portions of a cooperative frame in order to maximize the diversity gain at each multiplexing gain and we avoid the cooperation whenever it reduces the diversity gain compared to the case that source transmits alone. In difference with \cite{Krishnakumar-Naveen-Kumar}, we consider the more practical amplify-and-forward (AF) and decode-and-forward (DF) types protocols with real (not integer) asynchronous delays and examine the effect of the asynchronism on the DMT of the system from both the theoretical and the practical points of views.

The transmitter nodes send PAM signals in which information symbols are linearly modulated by a shaping waveform to be sent to the destination. We consider two different cases with respect to the length of the shaping waveforms used in the structure of the PAM signals. In case that the shaping waveforms have an infinite time-support, for example when the ``sinc'' waveform is used, the communication is carried out over a strictly limited bandwidth and it is shown that asynchronism does not affect the DMT performance of the system. However, when the shaping waveforms have a limited time-support which is in fact the case in practice, the transmitted signals in the frequency domain lie in a spectral mask which does not have a limited support. Although the tails of the spectrum are usually neglected because they are below the noise level, they may expand the bandwidth when the system is analyzed at high values of SNR. In this case, it is argued that
\begin{itemize}
\item both the OSDF and the NSDF protocols provide better diversity gains throughout the range of the multiplexing gain over the asynchronous network compared to those of the corresponding synchronous networks. In addition, similar to what is reported in \cite{Shuangqing_Wei}, the equivalent channel model in high values of SNR becomes the same as that of a parallel channel with the number of independent links equal to the number of transmitting nodes.
\item the NAF protocol provides a better diversity gain in the asynchronous scenario compared to the synchronous scenario throughout the range of the multiplexing gain. In particular, this protocol results in the same DMT as that of the $2\times 1$ multiple-input single-output (MISO) channel in a single relay asynchronous cooperative network.
\item the OAF protocol provides the same diversity gain over both asynchronous and the corresponding synchronous networks for all multiplexing gains.
\end{itemize}
The rest of the paper is organized as follows. In Section \ref{Section_Asyn_RelayNetwok}, the underlying asynchronous relay network is discussed and the system model is presented. The DMT analysis of the asynchronous NSDF, OSDF, NAF, and OAF protocols are detailed respectively in Sections \ref{Section_NSDF}, \ref{Section_OSDF}, \ref{Section_NAF}, and \ref{Section_OAF}. For each protocol, the DMT performance is analyzed for both cases of having infinite and finite length shaping waveforms. This paper is discussed and concluded in Section \ref{Discussion}.

\section{Asynchronous Relay Networks}\label{Section_Asyn_RelayNetwok}
\subsection{Notations and Definitions}
In this work, letters with underline, $\underline{x},\underline{X}$, denote vectors, and boldface uppercase letters, $\mathbf{X}$, denote matrices. The superscripts $(\cdot)^T$, $(\cdot)^*$, and $(\cdot)^\dagger$ denote the transpose,
conjugate, and conjugate transpose of the corresponding vector or matrix, respectively. $\mathbf{I}_n$ is the identity matrix of dimension $n$. $(x)^+$ denotes $\max\{0,x\}$. $\doteq$ is used to show the exponential equality. For example, $f(\rho) \doteq \rho^b$ if $\lim_{\rho \rightarrow \infty} \frac{\log{f(\rho)}}{\log{\rho}} = b$.

For a family of variable rate codes $\{\mathcal{C}(\rho)\}$ with signal to noise ratio (SNR), $\rho$, the multiplexing gain $r$ and the diversity gain $d(r)$ are defined as
\begin{equation}
  \lim_{\rho \rightarrow \infty} \frac{R(\rho)}{\log{\rho}} \triangleq r,~~~~\lim_{\rho \rightarrow \infty} \frac{\log P_e(\rho)}{\log{\rho}} \triangleq -d(r),
\end{equation}
where $R(\rho)$ is the transmission rate and $P_e(\rho)$ is the average error probability of the code $\mathcal{C}(\rho)$. It is shown in \cite{Zheng-Tse} that there is a tradeoff between $r$ and $d(r)$ known as the diversity multiplexing tradeoff (DMT). Moreover, for each multiplexing gain $r$,
\begin{equation}
  d(r) \le d^*(r),
\end{equation}
where $d^*(r)$ is the outage diversity which is defined as the negative exponent of $\rho$ in the outage probability expression $P_\mathcal{O}(R(\rho)) \doteq \rho^{-d^*(r)}$.

\subsection{System Description}\label{SYSTEM_MODEL}
\begin{figure}
\centering
\includegraphics[width=5cm, height = 4cm]{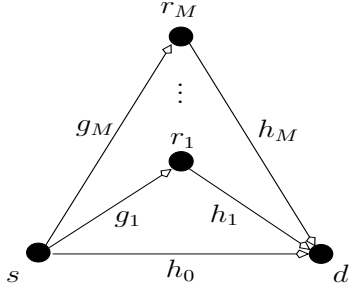}
\caption{System structure}\label{System_Model_Figure}
\end{figure}

We consider a network containing one source node, one destination node, and $M$ parallel relay nodes as shown in Fig. \ref{System_Model_Figure}. $h_i$ and $g_i$ are fading coefficients representing the links from the $i$-th transmitting node to the destination and from the source to the $i$-th relay, respectively. All channel gains are assumed to be independent and identically distributed (i.i.d.) complex Gaussian random variables with zero mean and unit variance $\mathbb{C}\mathcal{N}(0,1)$. They are constant within the transmission of a frame and vary independently at the beginning of each frame.

We assume half-duplex signal transmission whereby each node can either transmit or receive but not both at any given time instant. Communication between the source and the
destination is carried out in two phases. First, the source broadcasts its message to the relays and the destination in $p$ channel uses. Second, the relays retransmit it to the destination in $q$ channel uses based on the DF or the AF types relaying protocols. In the former, only those relays that are not in outage independently re-encode the source message and resend it to the destination; however in the latter, all relays perform linear transformations over the received signal and retransmit it to the destination. Assuming $\ell$ is the length of a cooperative frame, $\ell=p+q$.
We consider both cases of non-orthogonal and orthogonal cooperating protocols where in the second phase of the former the source sends a new codeword of length $q$, while in the latter, the source becomes silent in the second phase. For each protocol, the case that the source transmits alone over a fix portion of a frame equal to $p/\ell$ for all multiplexing gains is considered first. Then, $\kappa \triangleq p/q$ is optimized to maximize the diversity gain at each multiplexing gain. Since the source may transmit over both phases, it may have two independent codebooks of proper codewords' length. The cooperation is avoided whenever it reduces the diversity gain compared to the case that the source transmits alone. Each node knows the channel state information (CSI) of its incoming links. The destination knows the CSI of all the links, the number of the helping nodes, and their corresponding asynchronous delays.

\subsubsection*{Phase I}
By assuming that the source uses an i.i.d. Gaussian codebook with codewords of length $p$ in the first phase, its transmitted signal is given by
\begin{equation}
  x'_0(t) = \sum_{k=0}^{p-1}x'_0(k)\psi_0(t-kT_s),
\end{equation}
where $\underline{x}'_0 = \left[x'_0(0), x'_0(1), \ldots, x'_0(p-1)\right]^T$ is the transmitted codeword corresponding to the source message, $T_s$ is the symbol interval, and $\psi_0(t)$ is a unit energy shaping waveform with non-zero duration $uT_s$ over $t \in [0,uT_s]$ for a positive integer value of $u$.  $\psi_0(t)$ can simply be the shifted version of the truncation of a well-designed waveform in the interval $[-uT_s/2,uT_s/2]$ to the right by $uT_s/2$. The received signals in the first phase at the destination and the $i$-th relay $(i=1,\ldots, M)$, respectively, are modeled by
\begin{eqnarray}
  y_d(t) &=& h_0x'_0(t) + z_d(t),\\
  y_{r_i}(t) &=& g_ix'_0(t) + z_{r_i}(t),
\end{eqnarray}
where $z_d(t)$ and $z_i(t)$ are additive white noises modeled by complex Gaussian random variables $\mathbb{C}\mathcal{N}(0,\sigma^2_d)$ and $\mathbb{C}\mathcal{N}(0,\sigma^2_r)$, respectively.

\subsubsection*{Phase II}
Let $\mathcal{D}$ be a set containing index of the nodes participating in the second phase. Clearly, for the AF type protocols $\mathcal{D}$ contains index of all the relays;  however, for the DF type protocols it contains only index of the relays that can fully decode the source message. $\mathcal{D}$ contains index of the source which is zero in non-orthogonal protocols. In the DF type protocols, each relay is supported by an independent identically distributed (i.i.d.) random Gaussian codebook with codewords of length $q$. In the AF type protocols, the received signals at the relays are linearly processed and retransmitted to the destination. In both cases, the $i$-th relay uses a unit energy shaping waveform $\psi_i(t)$ with non-zero duration $uT_s$ to transmit its message.

The $i$-th transmitted signal at the second phase is received at the destination by $\tau_i$ second asynchronous delay with reference to the earliest received signal. Without loss of generality, in non-orthogonal protocols, we assume that the source signal is the earliest received signal at the destination and the delays of the other received signals are measured with reference to this signal; hence, $\tau_0=0$. In orthogonal protocols, we assume that $\tau_1=0$. In any case, if $m$ relays participates in the second phase, we index the nodes such that $\tau_0 < \tau_1 < \tau_2 < \ldots < \tau_m$. Since the relative delays are due to the random nature of the medium, the probability of the event that two of them are equal is zero. In this work, we assume that $\tau_i$ is less than a symbol interval. Generalizing the results to the case that asynchronous delays can be greater than a symbol interval is straightforward. Let $x_i(t)$ be the transmitted signal by the $i$-th transmitting node, $i \in \mathcal{D}$. The received signal at the destination in the second phase is modeled by
\begin{equation}
  y_d(t) = \sum_{i \in \mathcal{D}}h_ix_i(t-\tau_i) + z_d(t).
\end{equation}

\subsection{Discrete System Model}
Let $E_m$ be the event of any $m$ relays participate in the second phase. $E_0$ corresponds to the case that only the source transmits in the second phase. Assume $E_m$ occurs, $0 < m \le M$. $\mathcal{D} = \{0,1,2,\ldots,m\}$ is the index set pointing out to participating nodes in the second phase. Without loss of generality, we assume that $0 = \tau_0 < \tau_1 < \tau_2 < \ldots < \tau_m$. Note that for AF type protocols, $m=M$. To acquire the sufficient statistic of the received signal, it is passed through a set of parallel filters each of them matched on one of the incoming links \cite{Verdu}. The output of the $i$-th matched filter $i\in \{0,1,\ldots,m\}$ sampled at $t = (k+1)T_s+\tau_{i}, ~ k=0,\ldots, q-1$, is given by
\begin{align}\nonumber
 y_{d,i}(k) &= \int_{kT_s + \tau_i}^{(k+u)T_s + \tau_i}y_d(t)\psi^*_i(t-kT_s-\tau_i)dt\\
 &= \sum_{j \in \mathcal{D}} h_{i,j}\sum_{n=-u}^{u} \gamma_{i,j}(n) x_j(k+n) + z_{d,i}(k)
\end{align}
where $x_j(n)=0,\, \forall\, n<0$,
\begin{align*}
  \gamma_{i,j}(n) &= \int_0^{uT_s}\psi_j(t-nT_s+\tau_{i,j})\psi_i^*(t)dt,\\
  z_{d,i}(k) &= \int_{kT_s+\tau_i}^{(k+u)T_s+\tau_i}z_d(t)\psi^*_i(t-kT_s-\tau_{i})dt,
\end{align*}
and the relative delay $\tau_{i,j}$ is defined as
\begin{equation}\label{tau_ij}
  \tau_{i,j} \triangleq \tau_i -\tau_j.
\end{equation}
Since the shaping waveforms are of length $u$ symbol intervals and the relays are asynchronous, every transmitted symbol of a relay is interfered by $2(u-1)$ symbols (if not zero) of the same transmitted stream and $2u-1$ symbols (if not zero) of every transmitted stream by other relays. This can be verified by checking that, $\gamma_{i,i}(u) = \gamma_{i,i}(-u) = 0,\, \forall\, i\in\mathcal{D}$. Moreover, for $j\ne i$ if $\tau_{i,j}<0$, then $\gamma_{i,j}(u)=0$. Else if $\tau_{i,j}> 0$, then $\gamma_{i,j}(-u)=0$. The received signal vector at the output of the $i$-th matched filter is given by
\begin{equation}\label{Eqn_y_di_finite_u}
  \underline{y}_{d,i} = \sum_{j \in \mathcal{D}} h_j\mathbf{\Gamma}_{i,j}\underline{x}_j+\underline{z}_i,
\end{equation}
where
\begin{align*}
  \underline{y}_{d,i} &= [y_{d,i}(0),y_{d,i}(1),\ldots,y_{d,i}(q-1)]^T,\\
  \underline{x}_j &= [x_j(0),x_j(1),\ldots,x_j(q-1)]^T,\\
  \underline{z}_{d,i} &= [z_{d,i}(0),z_{d,i}(1),\ldots,z_{d,i}(q-1)]^T,
\end{align*}
and $\mathbf{\Gamma}_{i,j}$ is given in \eqref{Eqn_Gamma_ij} in the general form; however, $\gamma_{i,j}(-u)$ or $\gamma_{i,j}(u)$ might be zero depending on $i,j$.
\begin{equation}\label{Eqn_Gamma_ij}
  \mathbf{\Gamma}_{i,j} = \left[
  \begin{array}{ccccccccccc}
    \gamma_{i,j}(0) & \gamma_{i,j}(-1)\cdots & \gamma_{i,j}(-u) &0 & 0 & \ldots & 0 &  \ldots  & 0\\
    \gamma_{i,j}(1) & \cdots & \gamma_{i,j}(-u+1) & \gamma_{i,j}(-u)& 0 & \ldots & 0 & \ldots & 0\\
    \ddots & \ddots & \ddots & \ddots & \ddots & \ddots & \ddots & \ddots & \ddots\\
    0 & \ldots & 0 & 0& \gamma_{i,j}(u) & \cdots & \gamma_{i,j}(0) & \cdots & \gamma_{i,j}(-u)\\
     \ddots & \ddots & \ddots & \ddots & \ddots & \ddots & \ddots & \ddots & \ddots\\
   0 &  0 &\cdots & 0 & \cdots & 0 & \gamma_{i,j}(u) & \cdots & \gamma_{i,j}(0)
  \end{array}\right].
\end{equation}
$\underline{z}_i$ is the colored noise vector with the covariance matrix given by
\begin{equation}
  \mathbf{\Phi}_{i,j} = \sigma_d^2\mathbf{\Gamma}_{i,j}.
\end{equation}

The output vectors of the matched filters at the second phase can be written in a long vector form as
\begin{equation}\label{MIMO_Channel_Model}
  \underline{y} = \mathbf{H}\underline{x}+\underline{z},
\end{equation}
where
\begin{align}\nonumber
  \underline{x} &= \left[\underline{x}_0^T, \underline{x}_1^T, \ldots, \underline{x}_m^T\right]^T,\\ \nonumber
  \underline{y} &= \left[\underline{y}_{d,0}^T, \underline{y}_{d,1}, \ldots, \underline{y}_{d,m}\right]^T,\\ \nonumber
  \underline{z} &= \left[\underline{z}_{d,0}^T, \underline{z}_{d,1}, \ldots, \underline{z}_{d,m}\right]^T,\\ \label{H-Model-NSDF-mRelay}
  \mathbf{H} &= \mathbf{\Xi}(\mathbf{I}_q\otimes\hat{\mathbf{H}}),
\end{align}
and
\begin{align}\nonumber
\hat{\mathbf{H}} &= \texttt{diag}\{h_0,h_1,\ldots,h_m\},\\
\mathbf{\Xi} &= \left[
\begin{array}{ccccc}
 \mathbf{\Gamma}_{0,0} & \mathbf{\Gamma}_{0,1} & \mathbf{\Gamma}_{0,2} & \ldots & \mathbf{\Gamma}_{0,m}\\
 \mathbf{\Gamma}_{1,0} & \mathbf{\Gamma}_{1,1} & \mathbf{\Gamma}_{1,2} &\ldots & \mathbf{\Gamma}_{1,m}\\
 \vdots & \vdots & \vdots &  & \vdots\\
 \mathbf{\Gamma}_{m,0} & \mathbf{\Gamma}_{m,1} & \mathbf{\Gamma}_{m,2} & \ldots & \mathbf{\Gamma}_{m,m}
\end{array}\right].
\end{align}
Equation (\ref{MIMO_Channel_Model}) represents a simple multiple-input multiple-output (MIMO) channel model with correlated noise vector $\underline{z}$. The covariance matrix of $\underline{z}$ is given by
\begin{eqnarray}\label{Phi-Model-NSDF-mRelay}
  \mathbf{\Phi} = \sigma_d^2~\mathbf{\Xi}.
\end{eqnarray}
One can check that $\gamma_{i,j}(n) = \gamma^*_{j,i}(-n),\, n=0,1,\ldots,q-1$. Hence, $\mathbf{\Gamma}_{i,j} = \mathbf{\Gamma}_{j,i}^\dagger$ and $\mathbf{\Xi}$ is a Hermitian matrix with banded Toeplitz blocks of order $u$.

\subsection{Properties of Matrix $\mathbf{\Xi}$}
For an absolutely summable infinite complex sequence $\{\gamma_{i,j}(k),\,k\in \mathbb{Z}\}$, where $\mathbb{Z}$ is the set of integers, the $2\pi$-periodic Discrete-Time-Fourier-Transform (DTFT) is defined as \cite{Oppenheim-Schafer}
\begin{equation}\label{Eqn_DTFT}
  \Gamma_{i,j}(\omega) \triangleq \sum_k \gamma_{i,j}(k) e^{-\xi\omega k},~ \omega \in [0,2\pi],
\end{equation}
where $\xi = \sqrt{-1}$. Define Matrix $\mathbf{\Gamma}(\omega)$ as
\begin{equation}\label{Eqn_Gamma_ij_Omega}
  \mathbf{\Gamma}(\omega) \triangleq \left[
  \begin{array}{cccc}
    \Gamma_{0,0}(\omega) & \Gamma_{0,1}(\omega) & \cdots & \Gamma_{0,m}(\omega)\\
    \Gamma_{1,0}(\omega) & \Gamma_{1,1}(\omega) & \cdots & \Gamma_{1,m}(\omega)\\
    \vdots & \vdots & \cdots & \vdots\\
    \Gamma_{m,0}(\omega) & \Gamma_{m,1}(\omega) & \cdots & \Gamma_{m,m}(\omega)\\
  \end{array}\right].
\end{equation}
$\mathbf{\Gamma}(\omega)$ is a Hermitian matrix, i.e., $\mathbf{\Gamma}(\omega) = \mathbf{\Gamma}(\omega)^\dagger$. In the sequel, we will need the following theorem from \cite{Gazzah-Regalia-Delmas}.
\begin{theorem}\label{Theorem_Szego_BlockToeplitz}
  Let $\lambda_k,\, k = 1,2,\ldots, (m+1)q$, be the $k$-th eigenvalue of $\mathbf{\Xi}$. Let $\mu_k(\omega),\, k = 1,2,\ldots, m+1$, be the $k$-the eigenvalue of $\mathbf{\Gamma}(\omega)$. For all continuous functions, $F(\cdot)$, one has
\begin{equation}\nonumber
  \lim_{q \to \infty} \frac{1}{q}\sum_{k=1}^{(m+1)q}F(\lambda_k) = \frac{1}{2\pi}\int_{-\pi}^\pi\sum_{k=1}^{m+1} F(\mu_k(\omega))d\omega.
\end{equation}
Moreover the eigenvalues of $\mathbf{\Xi}$ lie in $[\min_{k,\omega} \mu_k(\omega),\max_{k,\omega}\mu_k(\omega)]$ and if they are sorted in a descending order, then for every positive integer $a$, the lowest (largest) $a$ eigenvalues of $\mathbf{\Xi}$ are convergent in $q$, i.e.,
  \begin{align*}
    \lim_{q\to\infty} \lambda_{(m+1)q-a+1} &= \min_{k,\omega} \mu_k(\omega)\\ \lim_{q\to\infty} \lambda_a &= \max_{k,\omega} \mu_k(\omega).
  \end{align*}
\end{theorem}

See \cite{Gazzah-Regalia-Delmas} for the proof. The above theorem extends the results of the Szeg\"{o}'s Theorem in \cite{Grenander-Szego} to Hermitian block Toeplitz matrices.

\begin{lemma}\label{Lemma_Xi_Expression}
Matrix $\mathbf{\Gamma}(\omega)$ can be expressed as
  \begin{equation}\nonumber
    \mathbf{\Gamma}(\omega) = \int_0^{T_s} \left(\sum_{i=0}^u \underline{\psi}(t+iT_s)
    e^{\xi\omega i}\right)^\dagger \sum_{i=0}^u\underline{\psi}(t+iT_s)e^{\xi\omega i} dt.
  \end{equation}
where $\underline{\psi}(t) \triangleq [\psi_0(t),\psi_1(t-\tau_{1,0}), \ldots,\psi_m(t-\tau_{m,0})]$.
\end{lemma}

The proof is given in Appendix \ref{Apx_Proof_Lemma_Xi_Expression}.

\begin{proposition}\label{Prop_det_Xi_k}
$\mathbf{\Gamma}(\omega)$ is a semi-positive definite matrix $\forall\, \omega\in[0,2\pi]$, i.e., $\det\mathbf{\Gamma}(\omega)\ge 0$. The equality holds if and only if $\exists ~\underline{c} \in \mathbb{C}^{(m+1) \times 1}, \exists~ \omega \in [0,2\pi]$ such that
\begin{eqnarray}\label{gis_condition}
  \left(\sum_{i=0}^u\underline{\psi}(t+iTs)e^{\xi\omega i}\right)\underline{c} = 0,~~~ \forall t \in [0,T_s],
\end{eqnarray}
where $\mathbb{C}$ is the field of complex numbers.
\end{proposition}
\begin{proof}
Proving that $\mathbf{\Gamma}(\omega)$ is a semi-positive definite matrix $\forall \,\omega$ is a direct result of Lemma \ref{Lemma_Xi_Expression}. Hence,
\begin{equation}\nonumber
  \forall \underline{c} \in \mathbb{C}^{(m+1) \times 1}, ~~\underline{c}^\dagger \mathbf{\Gamma}(\omega) \underline{c} \ge 0.
\end{equation}
Using Lemma \ref{Lemma_Xi_Expression}, $\underline{c}^\dagger \mathbf{\Gamma}(\omega) \underline{c}$ is equal to
\begin{align*}
\underline{c}^\dagger \mathbf{\Gamma}(\omega) \underline{c} =
  \int_0^{T_s}\left[\left(\sum_{i=0}^u\underline{\psi}(t+iT_s)
  e^{\xi\omega i}\right)\underline{c}\right]^\dagger \left(\sum_{i=0}^u\underline{\psi}(t+iT_s)
  e^{\xi\omega i}\right)\underline{c}\, dt.
\end{align*}
If $\underline{c}^\dagger \mathbf{\Gamma}(\omega) \underline{c} =
0$, there must exist $\underline{c} \in \mathbb{C}^{(m+1) \times
1}$ such that
\begin{equation}\nonumber
  \left(\sum_{i=0}^u\underline{\psi}(t+iT_s)e^{\xi\omega i}\right)\underline{c} = 0,
   ~~~ \forall t\in[0,T_s].
\end{equation}
This concludes the proof.
\end{proof}

According to Proposition \ref{Prop_det_Xi_k}, if the shaping waveforms do not satisfy in (\ref{gis_condition}), $\mathbf{\Gamma}(\omega)$ is a positive definite matrix, and it has $(m+1)$ non-zero positive real eigenvalues. Since all the $\{\gamma_{i,j}(k)\}$ sequences are assumed to be absolutely summable, $\sum_{i=1}^{m+1} \mu_i(\omega)$ which is equal to the trace of $\mathbf{\Gamma}(\omega)$ is a bounded value. Consequently, all eigenvalues of $\mathbf{\Gamma}(\omega)$ are also bounded. In this case, where according to Theorem \ref{Theorem_Szego_BlockToeplitz}, $\mathbf{\Xi}$ is a full-rank matrix with all bounded real eigenvalues, the discrete system model presented in \eqref{MIMO_Channel_Model} is used.

\subsubsection{When $u$ is finite}
$\sum_{i=0}^{u}\underline{\psi}(t+iT_s)e^{\xi\omega i}$ is a vector containing the DTFT of the samples of the vector $\underline{\psi}(t')$ at $t'=t+iT_s,\, i\in \mathbb{Z},\, \forall\, t\in[0,T_s]$. For a finite value of $u$, the spectrum of the waveforms has infinite support and occupies the whole frequency axis. Hence, the signal cannot be recovered from its samples and the DTFT of a set of samples (for a specific $t \in [0,T_s]$) is a function of the shift $t$ and does not necessarily relate to the DTFT of another set of samples. Hence, equation \eqref{gis_condition} does not hold almost always when $u$ is a finite value.

\subsubsection{When $u \to \infty$}
For an even value of $u$, define $\underline{\hat{\psi}}(t) \triangleq \underline{\psi}(t+\frac{u}{2}T_s)$. Hence, $\sum_{i=0}^u\underline{\psi}(t+iT_s)e^{\xi\omega i} = e^{\xi\omega u/2} \sum_{i=-u/2}^{u/2}\underline{\hat{\psi}}(t+iT_s)e^{\xi\omega i}$ and
equation \eqref{gis_condition} can be rewritten based on $\underline{\hat{\psi}}(t)$ as follows.
\begin{equation}\label{Eqn_gis_condition2}
  \left(\sum_{i=-u/2}^{u/2}\hat{\underline{\psi}}(t+iT_s)e^{\xi\omega i}\right)\underline{c} = 0, ~~~ \forall t\in[0,T_s].
\end{equation}
Let $u \to \infty$. In this case, the communication is carried out over a strictly limited bandwidth $W$ and $\lim_{u\to \infty} \sum_{i=-u/2}^{u/2}\\\underline{\hat{\psi}}(t+iT_s)e^{\xi\omega i}$ is a vector containing the DTFT of the elements of the vector $\underline{\hat{\psi}}(t')$ sampled at $t'=t+iT_s,\, i\in \mathbb{Z},\, \forall\, t\in[0,T_s]$. If the frequency bandwidth $W$ is such that $W \le \frac{1}{2T_s}$, then the shift property of the DTFT for non-integer delays is held (see Appendix \ref{Apx_Shift_Property_of_DFT}) and equation \eqref{Eqn_gis_condition2} can be written as follows.
\begin{equation}\label{Eqn_c_Choices}
  e^{-\xi\omega t}\sum_{i=0}^m c_i\hat{\Psi}_j(-\omega)e^{\xi\omega\tau_{i,0}} = 0, ~~~\forall\, t \in [0,T_s],
\end{equation}
where $\hat{\Psi}_j(\omega)$ is the DTFT of the samples of $\hat{\psi}_j(t')=\psi_j(t'+uT_s/2)$. It is obvious that for each $\omega \in [0,2\pi]$, there are many choices for vector $\underline{c}$ which satisfy equation \eqref{Eqn_c_Choices}. This is because the exponential term containing the shift parameter $t$ appears as the multiplicative factor of all the coefficients, $c_i$'s, and does not affect the roots of this equation. Therefore,  $\mathbf{\Gamma}(\omega)$ is not full rank which according to Theorem \ref{Theorem_Szego_BlockToeplitz} implies that $\mathbf{\Xi}$  is not full-rank either (for large values of $q$).

To determine the rank order of $\mathbf{\Gamma}(\omega)$ in this case, One can see that when $u\to \infty$
\begin{equation}
  \Gamma_{i,j}(\omega) = \hat{\Psi}_j(-\omega)e^{\xi\omega\hat{\tau}_{i,j}}\hat{\Psi}^*_i(\omega).
\end{equation}
Hence, $\mathbf{\Gamma}(\omega)$ in \eqref{Eqn_Gamma_ij_Omega} can be re-written in this case as
\begin{equation}
  \mathbf{\Gamma}(\omega) = \mathbf{\Psi}(-\omega)\mathbf{E}(\underline{1}\otimes\underline{e}^T)
  \mathbf{\Psi}^*(\omega),
\end{equation}
where $\underline{1}$ is a vector of length $m+1$ with all entries equal to one, $\otimes$ is the Kronecker product, and
\begin{align*}
  \mathbf{\Psi}(\omega) &= \texttt{diag}\{\Psi_0(\omega),\Psi_1(\omega),\ldots,\Psi_m(\omega)\}\\
  \mathbf{E} &= \texttt{diag}\{e^{\xi\omega\hat{\tau}_{0,0}},e^{\xi\omega\hat{\tau}_{1,0}},
  \ldots,e^{\xi\omega\hat{\tau}_{m,0}}\}\\
  \underline{e} &= [1,e^{\xi\omega\hat{\tau}_{0,1}},\ldots,e^{\xi\omega\hat{\tau}_{0,m}}]^T.
\end{align*}
As can be seen, all rows of $\mathbf{\Gamma}_{i,j}$ are linearly dependent in this case and, therefore, it has rank order one. In this case, one matched filter is adequate to acquire the sufficient statistic. However, since the received signal is strictly bandwidth limited,  sampling with $f_s = 2W$ (without matched filtering) is enough for this purpose. The discrete model of the channel in this case, which is used throughout the paper when $u\to \infty$, is given as follows.
\begin{align}\label{Eqn-yd-infiniteu}
  \underline{y}_d &= \sum_{j=0}^m h_j\mathbf{\Gamma}_j\underline{x}_j+\underline{z}_d,
\end{align}
where $\underline{z}_d$ is the white Gaussian noise vector with covariance matrix $\sigma_d^2\mathbf{I}_q$. Assuming $\gamma_j(k) = \psi_j(kT_s-\tau_{j,0}), \, k = -q+1,\ldots,0,\ldots,q-1$, is the $k$-th sample of the shaping waveform, $\mathbf{\Gamma}_j$ is given by

\begin{equation}\label{Eqn_Gamma_j}
  \mathbf{\Gamma}_j = \left[
  \begin{array}{cccc}
    \gamma_j(0) & \gamma_j(-1) & \cdots & \gamma_j(-q+1)\\
    \gamma_j(1) & \gamma_j(0) & \cdots & \gamma_j(-q+2)\\
    \ddots & \ddots & \cdots & \ddots\\
    \gamma_j(q-1) & \gamma_j(q-2) & \cdots & \gamma_j(0)
  \end{array}\right].
\end{equation}
\begin{proposition}\label{Prop_RankofGammaj}
For well-designed shaping waveforms with non-zero spectrum over the bandwidth $W$ and the sampling frequency $f_s = 2W$, $\mathbf{\Gamma}_j$ is a full rank matrix $\forall\,q < \infty$ with all bounded eigenvalues.
\end{proposition}

The proof is given in Appendix \ref{Appendix_Proof_Prop_RankofGammaj}.

\section{Asynchronous NSDF Relaying Protocol}\label{Section_NSDF}
For our DF type protocols, the outage probability, $P_\mathcal{O}$, is calculated as follows.
\begin{equation}\label{Outage_Prob}
  P_\mathcal{O} =  \sum_{m=0}^MPr(I_{E_m}<R)Pr(E_m),
\end{equation}
where $I_{E_m}$ is the mutual information between the source
and the destination when $E_m$ occurs. Let $\mathcal{D}$ be the index set corresponding to the event
$E_m$. For a transmission rate $R$, the probability of the occurrence of the event $E_m$, $Pr(E_m)$, is given by
\begin{align*}
  & Pr(E_m) = \prod_{k \in \mathcal{D}}Pr(I_{s,r_k} \ge R)\prod_{k \not\in \mathcal{D}}Pr(I_{s,r_k} < R)\\
  &= \prod_{k \in \mathcal{D}}Pr\left(p\log(1+\rho|g_k|^2)\ge \ell R\right) \prod_{k \not\in \mathcal{D}}Pr\left(p\log(1+\rho|g_k|^2) < \ell R\right)\\
  &= \prod_{k \in \mathcal{D}} Pr\left(|g_k|^2 \ge \frac{2^{\frac{\ell R}{p}}-1}{\rho}\right)\prod_{k  \not \in \mathcal{D}} Pr\left(|g_k|^2 < \frac{2^{\frac{\ell R}{p}}-1}{\rho}\right)\\
  &= \prod_{k \in \mathcal{D}} e^{-\frac{2^{\frac{\ell R}{p}}-1}{\rho}}
  \prod_{k \not \in \mathcal{D}}\left(1-e^{-\frac{2^{\frac{\ell R}{p}}-1}{\rho}}\right),~~~
\end{align*}
where $I_{s,r_k}$ is the mutual information between the source
and the $k$-th relay in the first phase. The last equality
comes from the fact that $|g_k|^2$ has exponential distribution
with parameter $\lambda_k=1$. By considering $R = r\log \rho$ for large values of $\rho$,
\begin{align*}
  e^{-\frac{2^{\frac{\ell R}{p}}-1}{\rho}} &= e^{-\frac{\rho^{\frac{\ell r}{p}}-1}{\rho}}
  \doteq \left\{
  \begin{array}{ll}
  1-\rho^{-\left(1-\frac{\ell r}{p}\right)}, & 0 \le r \le \frac{p}{\ell}\\
  0,  & \frac{p}{\ell} < r.
  \end{array}\right.
\end{align*}
Since the diversity gain is zero for $r > 1$, we only consider
the case that $0 \le r \le 1$. Despite the relays which are in
outage with probability one for $ r > \frac{p}{\ell}$, the
source node continues transmitting signal to the destination.
Hence, $Pr(E_0) = 1$ when $\frac{p}{\ell} < r \le 1$. Thus,
\begin{equation}\label{Pr(E_m)}
  Pr(E_m) \doteq \left\{
  \begin{array}{ll}
  \rho^{-(1-\frac{\ell r}{p})(M-m)}, & 0 \le r \le \frac{p}{\ell},\\
  0, & \frac{p}{\ell} < r \le 1, ~ 1 \le m \le M\\
  1, & \frac{p}{\ell} < r \le 1, ~  m=0.
  \end{array}\right.
\end{equation}

\subsection{Asynchronous NSDF with Infinite Length Waveforms}\label{Section_NSDF_infinite_u}
For the case that $u \to \infty$ and all the transmitters use the same shaping waveform, the system is modeled by equation \eqref{Eqn-yd-infiniteu}. By assuming a uniform power distribution among all the transmitting nodes, the mutual information between the source and the destination when $E_m$ occurs is given by
\begin{align}
  I_{E_m} =&  \frac{p}{\ell}\log\left(1+\rho|h_0|^2\right) +  \frac{1}{\ell} \log\det\Big(\mathbf{I}_q +\rho \sum_{j=0}^m |h_j|^2\mathbf{\Gamma}_j\mathbf{\Gamma}_j^\dagger\Big).
\end{align}
Since all Toeplitz matrices asymptotically commute, they are normal and are diagonalized on the same basis \cite{Robert-Gray}. Moreover, according to Proposition \ref{Prop_RankofGammaj}, for proper designed shaping waveforms, $\mathbf{\Gamma}_j$ is a full rank Toeplitz matrix with all non-zero eigenvalues bounded. Hence, for large values of $\rho$, we obtain
\begin{equation}
  I_{E_m} \doteq  \frac{p}{\ell}\log\left(1+\rho|h_0|^2\right) + \frac{q}{\ell}\log\Big(1+\rho\sum_{i=0}^m|h_i|^2\Big).
\end{equation}
As can be seen, $I_{E_m}$ in this case is the same as that of the corresponding synchronous network given in \cite{Elia-Vinodh-Anand-Kumar}. Hence, The DMT performance of both networks are the same.

Define $\alpha_i = -\frac{\log |h_i|^2}{\log \rho}$. Let $\alpha = \min_{i\ge 1}\alpha_i$. We obtain,
\begin{equation}
  I_{E_m} \doteq  \Big[\frac{p}{\ell}(1-\alpha_0)^+ + \frac{q}{\ell}(1-\alpha)^+\Big]\log \rho,
\end{equation}
where $(x)^+ = \max \{0,x\}$. By proceeding in the footsteps of \cite{Zheng-Tse}, the outage probability at high values of SNR when $E_m$ occurs is obtained as
\begin{align}\nonumber
  P_{\mathcal{O}\mid E_m} &= Pr\left(I_{E_m} < R \right)\\ \nonumber
  &= Pr\Big(p(1-\alpha_0)^+ + q(1-\alpha)^+ < \ell r\Big)\\ \nonumber
  &= \int_{\mathcal{R}_{E_m}}p(\alpha_0,\ldots,\alpha_m)d\alpha_0 \ldots d\alpha_m\\ \nonumber
  &\doteq \int_{\mathcal{R}_{E_m}}\rho^{-\sum_{j=0}^m\alpha_j}d\alpha_0 \ldots d\alpha_m\\
  &\doteq \rho^{-d_{E_m}(r)},
\end{align}
where $p(\alpha_0,\ldots, \alpha_m)$ is the joint probability density function of the parameters $\alpha_0,\ldots, \alpha_m$; $\mathcal{R}_{E_m} = \{(\alpha_0,\alpha) \mid
p(1-\alpha_0)^+ + q(1-\alpha)^+ < \ell r,~ \alpha_0, \alpha \ge 0\}$, and
\begin{equation}
 d_{E_m}(r) = \inf_{p(1-\alpha_0)^+ + q(1-\alpha)^+ < \ell r}
 \alpha_0+\alpha.
\end{equation}
By solving the above optimization problem and using \eqref{Outage_Prob} and \eqref{Pr(E_m)} we get
\begin{theorem}
  For $u \to \infty$, the DMT performance of the NSDF protocol over the underlying asynchronous relay network for a fix value of $\kappa = \frac{p}{q}$ is given as follows.

  Let $\kappa_M = \frac{1+\sqrt{1+4M^2}}{2M}$. If $1 \le \kappa \le \kappa_M$,
  \begin{equation}\nonumber
    d^*(r) = M\Big(1-\frac{\ell}{p}r\Big)^++(1-r), ~~~ 0 \le r \le 1,
  \end{equation}
  else, for $\kappa \ge \kappa_M$,
  \begin{equation}\nonumber
    d^*(r) = \left\{
    \begin{array}{ll}
      (M+1)\big(1-\frac{M\ell}{(M+1)q}r\big), & 0\le r \le \frac{q}{\ell}\\
      \frac{\ell}{p}(1-r), & \frac{q}{\ell} \le r \le \frac{(M+1)p-\ell}{(M-1)\ell+p}\\
      (M+1)\big(1-\frac{M\ell+p}{(M+1)p}r\big), & \frac{(M+1)p-\ell}{(M-1)\ell+p} \le r \le \frac{p}{\ell}\\
      1-r, & \frac{p}{\ell} \le r \le 1.
    \end{array}\right.
  \end{equation}
  When $\kappa$ varies to maximize the DMT at each multiplexing gain $r$, it is given by
  \begin{equation}\nonumber
    d^*(r) = \left\lbrace
    \begin{array}{ll}
      (M+1)\left(1-\frac{M(1+\kappa_M)}{M+1}\right), & 0 \le r < \frac{1}{1+\kappa_M}\\
      \frac{(M+1-r)(1-r)}{(M-1)r+1}, & \frac{1}{1+\kappa_M} \le r \le 1.
    \end{array}\right.
  \end{equation}
  The optimal value of $\kappa$ for a gain $r$ is given by
  \begin{equation}\nonumber
    \kappa = \left\lbrace
    \begin{array}{ll}
      \kappa_M, & 0 \le r < \frac{1}{1+\kappa_M}\\
      \frac{1+(M-1)r}{M(1-r)}, & \frac{1}{1+\kappa_M} \le r \le 1.
    \end{array}\right.
  \end{equation}
\end{theorem}

The proof is given in \cite{Elia-Vinodh-Anand-Kumar} and is omitted here for brevity.

\subsection{Asynchronous NSDF with Finite Length Waveforms}\label{Section_NSDF_finite_u}
For a finite value of $u$, the mutual information between the source and the destination when $E_m$ occurs is given by
\begin{align}\label{IDsmR}
  I_{E_m} =& \frac{p}{\ell}\log(1+\rho|h_0|^2) + \frac{1}{\ell} \log\det\left(\mathbf{I}_{(m+1)q}
  +\mathbf{\Phi}^{-1}\mathbf{H\Sigma}_x\mathbf{H}^\dagger\right),
\end{align}
where the first and the second terms on the right hand side of the above equation are the resulted mutual information between the transmitting nodes and the destination, respectively, in the first and in the second phases. $\mathbf{\Sigma}_x$ is the autocorrelation matrix of the input vector $\underline{x}$. For simplicity, we consider a uniform power allocation for all the transmitting nodes in the second phase. Define $\mathcal{A} \triangleq \mathbf{I}_{(m+1)q}+\mathbf{\Phi}^{-1}\mathbf{H}\mathbf{\Sigma}_x
\mathbf{H}^\dagger$. By substituting
(\ref{H-Model-NSDF-mRelay}) and (\ref{Phi-Model-NSDF-mRelay}) into (\ref{IDsmR}), we have
\begin{equation}\nonumber
  \det \mathcal{A} = \det \left(\mathbf{I}_{(m+1)q}+\rho (\mathbf{I}_q\otimes\hat{\mathbf{H}}\hat{\mathbf{H}}^\dagger)\mathbf{\Xi}\right).
\end{equation}
$\mathbf{\Xi}$ is a hermitian matrix and can be decomposed as $\mathbf{\Xi} = \mathbf{V\Lambda V}^\dagger$, where $\mathbf{V}$ is a unitary
matrix and $\mathbf{\Lambda}$ is a diagonal matrix containing eigenvalues of $\mathbf{\Xi}$ on its main diagonal. According to proposition \ref{Prop_det_Xi_k}, for well-designed shaping waveforms, $\mathbf{\Xi}$ is a positive definite matrix with all eigenvalues real and bounded. By replacing all the eigenvalues by the smallest one, say $\lambda$, we get
\begin{equation}\nonumber
  \det \mathcal{A} \ge \det \left(\mathbf{I}_{(m+1)q}+\rho \lambda \mathbf{I}_q\otimes\hat{\mathbf{H}}\hat{\mathbf{H}}^\dagger\right).
\end{equation}
Since $\lambda$ is a bounded value, this lower bound is tight when $\rho \to \infty$. In this case, the mutual information between the source and the destination at high values of SNR is given by
\begin{align}\nonumber
  I_{E_m} &\doteq \log(1+\rho|h_0|^2) + \frac{q}{\ell} \sum_{i=1}^m \log(1+\rho|h_i|^2)\\
  & \doteq \Big[(1-\alpha_0)^+ + \frac{q}{\ell}\sum_{i=1}^m(1-\alpha_i)^+ \Big] \log \rho.
\end{align}
As can be seen, the resulted mutual information among the transmitting nodes and the destination is similar to that of a parallel channel with $(m+1)$ independent links. By proceeding in the footsteps of \cite{Zheng-Tse}, $P_{\mathcal{O}\mid E_m}$ for a transmission rate $R = r\log \rho$ is calculated as follows.
\begin{equation}\nonumber \label{P_O_Em}
  P_{\mathcal{O}\mid E_m} = Pr\left(I_{E_m} < r\log\rho \right)
  \doteq \rho^{-d_{E_m}(r)}
\end{equation}
where for $\alpha_i \ge 0,\, i=0,\ldots,m,$
\begin{equation}
 d_{E_m}(r) = \inf_{(1-\alpha_0)^++\frac{q}{\ell}\sum_{i=1}^m(1-\alpha_i)^+ < r} \sum_{i=0}^m \alpha_i.
\end{equation}
By solving the above optimization problem for a fix value of $\kappa = \frac{p}{q}$ , we obtain
\begin{lemma}\label{Lemma_dEm_NSDF_FixK}
\begin{equation}\nonumber
  d_{E_m}(r) = \left\{
  \begin{array}{ccc}
    1+m-\frac{\ell}{q}r, & 0 \le r \le \frac{mq}{\ell},\\
    1+\frac{mq}{\ell}- r, & \frac{mq}{\ell} < r \le 1.
  \end{array}\right.
\end{equation}
Clearly, when $m \ge \kappa+1$, then $\frac{mq}{\ell} \ge 1$. Hence,
\begin{equation}\nonumber
  d_{E_m}(r) = 1+m-\frac{\ell}{q}r,~~~~ 0 \le r \le 1.
\end{equation}
\end{lemma}

The proof is given in Appendix \ref{Proof_Lemma_dEm_NSDF_FixK}. The following theorem treats the case where there is only one relay in the network.
\begin{proposition}\label{Proposition_NSDF_1Relay}
For a finite value of $u$, the DMT performance of the NSDF protocol over the single relay asynchronous cooperative network for a fix value of $\kappa \ge 1$ is as follows.

If $1 \le \kappa \le \hat{\kappa}$
\begin{equation}\nonumber
d^*(r)=\left\{
    \begin{array}{ll}
      (1-\frac{\ell}{p}r) + (1-r), & 0 \le r \le \frac{p}{\ell}\\
      1-r, &  \frac{p}{\ell} \le r \le 1,
    \end{array}\right.
\end{equation}
else, for $\kappa \ge \hat{\kappa}$
\begin{equation}\nonumber
d^*(r)=\left\{
    \begin{array}{ll}
       2(1-\frac{\ell}{2q}r), & 0 \le r \le \frac{q}{\ell}\\
       1+\frac{q}{\ell}-r, & \frac{q}{\ell} \le r \le \frac{p^2}{\ell^2}\\
       (1-\frac{\ell}{p}r) + (1-r), & \frac{p^2}{\ell^2} \le r \le \frac{p}{\ell}\\
       1-r, & \frac{p}{\ell} \le r \le 1,
    \end{array}\right.
\end{equation}
where $\hat{\kappa} = \frac{1+\sqrt{5}}{2}$. If $\kappa$ varies to maximize the diversity gain, we get
\begin{equation}\nonumber
d^*(r) = \left\{
\begin{array}{ll}
  [1-(1+\frac{1}{\hat{\kappa}})r]+(1-r), & 0 \le r \le \frac{1}{\hat{\kappa}+1}\\
  (1-\sqrt{r})+(1-r),&  \frac{1}{\hat{\kappa}+1} \le r \le 1.
\end{array}\right.
\end{equation}
The optimum $\kappa$ corresponding to each $r$ is given by
\begin{equation}\nonumber
 \kappa = \left\{
\begin{array}{ll}
 \hat{\kappa}, & 0 \le r \le \frac{1}{\hat{\kappa}+1} \\
 \frac{\sqrt{r}}{1-\sqrt{r}} , & \frac{1}{\hat{\kappa}+1} \le r \le 1.
\end{array}\right.
\end{equation}
\end{proposition}

The proof is given in Appendix \ref{Proof_Proposition_NSDF_1Relay}.
Since both $Pr(E_m)$ and $P_{\mathcal{O} \mid E_m}$ required in (\ref{Outage_Prob}) are known, calculating DMT in a general network with $M>1$ relays is straightforward.
However, it is easier if we assume that the DMT performance of a simpler network containing $(M-1)$ relays is known. Let $d_M^*(r)$ be the DMT of the NSDF protocol over an $M$ relay asynchronous cooperative network. The following theorem concludes the results in the general case.
\begin{theorem}\label{Theorem_NSDF_MRelay}
For a finite value of $u$, the DMT of the  NSDF protocol over a general two-hop asynchronous cooperative network with $M$ relays for a fix $\kappa \ge 1$ is as follows.\\
If $\kappa \le \frac{M+\sqrt{M^2+4M}}{2}$,
\begin{equation}\nonumber
  d_M^*(r) = \Big(1-\frac{\ell}{p}r\Big)+d^*_{M-1}(r), ~~~ 0 \le r \le \frac{p}{\ell}.
\end{equation}
Else, for $\kappa \ge \frac{M+\sqrt{M^2+4M}}{2}$
\begin{equation}\nonumber
d_M^*(r) = \left\{
\begin{array}{ll}
 (1-\frac{\ell}{p}r)+d^*_{M-1}(r), & 0 \le r \le \frac{Mq}{\ell}\\
 1+\frac{Mq}{\ell}-r, & \frac{Mq}{\ell} \le r \le \frac{p^2}{\ell^2}\\
 M(1-\frac{\ell}{p}r)+1-r, & \frac{p^2}{\ell^2} \le r \le \frac{p}{\ell},\\
 (1-r), & \frac{p}{\ell} \le r \le 1,
\end{array}\right.
\end{equation}
When $\kappa$ varies to maximize the diversity gain at each multiplexing gain $r$, we have
\begin{equation}\nonumber
  d(r) = \left\{
  \begin{array}{ll}
    M[1-(1+\frac{1}{\hat{\kappa}})r]+(1-r), & 0 \le r \le \frac{1}{1+\hat{\kappa}}\\
    M(1-\sqrt{r})+(1-r), & \frac{1}{1+\hat{\kappa}} \le r \le 1.
  \end{array}\right.
\end{equation}
where $\hat{\kappa} = \frac{1+\sqrt{5}}{2}$. The optimum $\kappa$ corresponding to each $r$ is given by
\begin{equation}\nonumber
  \kappa = \left\{
  \begin{array}{ll}
    \hat{\kappa}, & 0 \le r \le \frac{1}{1+\hat{\kappa}}\\
    \frac{\sqrt{r}}{1-\sqrt{r}}, & \frac{1}{1+\hat{\kappa}} \le r < 1.
  \end{array}\right.
\end{equation}
\end{theorem}

The proof is given in Appendix \ref{Proof_Theorem_NSDF_MRelay}.
\begin{figure}
\centering
\includegraphics[width=9cm, height = 7cm]{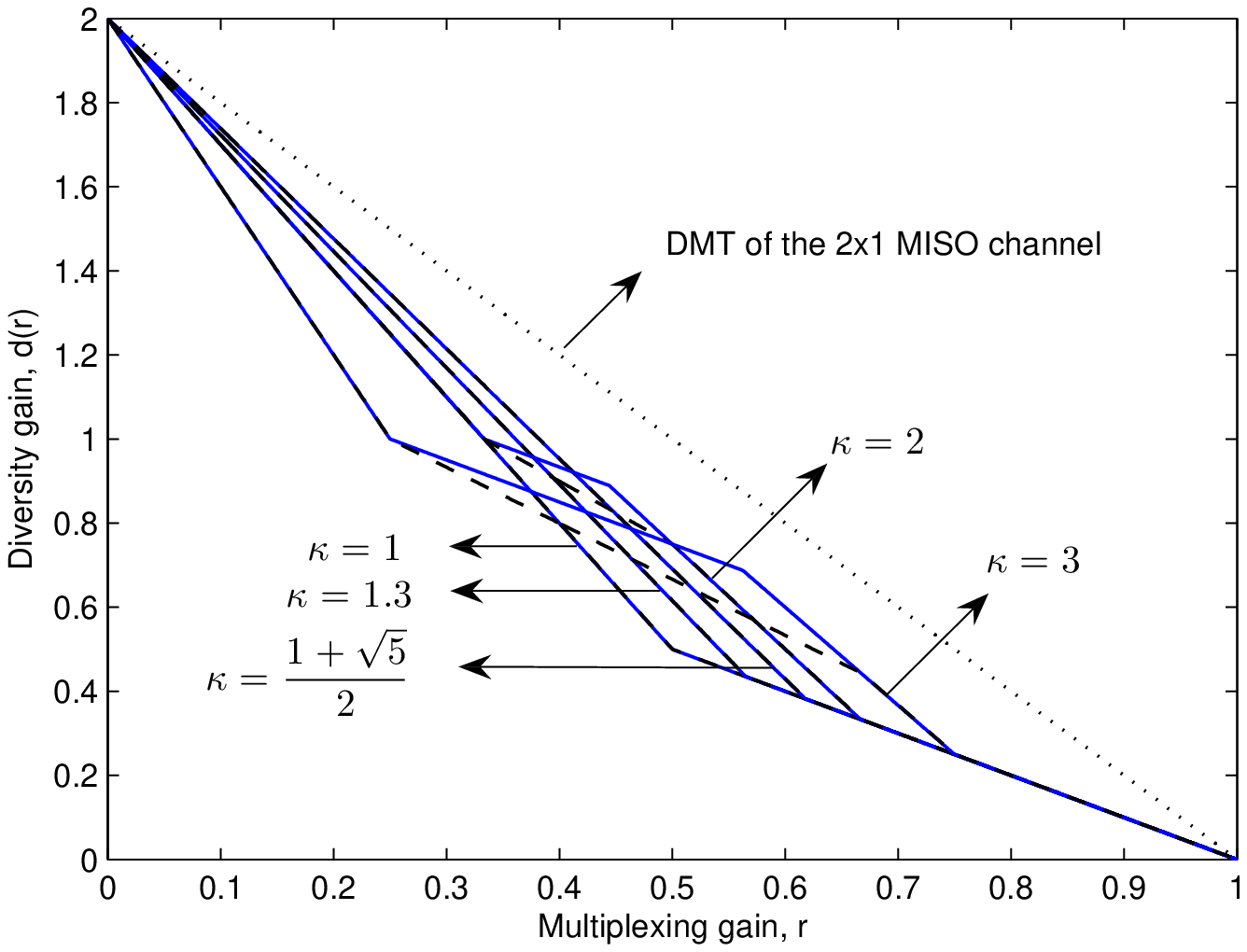}
\caption{The DMT performances of the asynchronous NSDF protocol over a single relay network for both finite and infinite length shaping waveforms and for various values of $\kappa > 1$.}\label{NSDF_VariousK_Figure}
\centering
\includegraphics[width=9cm, height = 7cm]{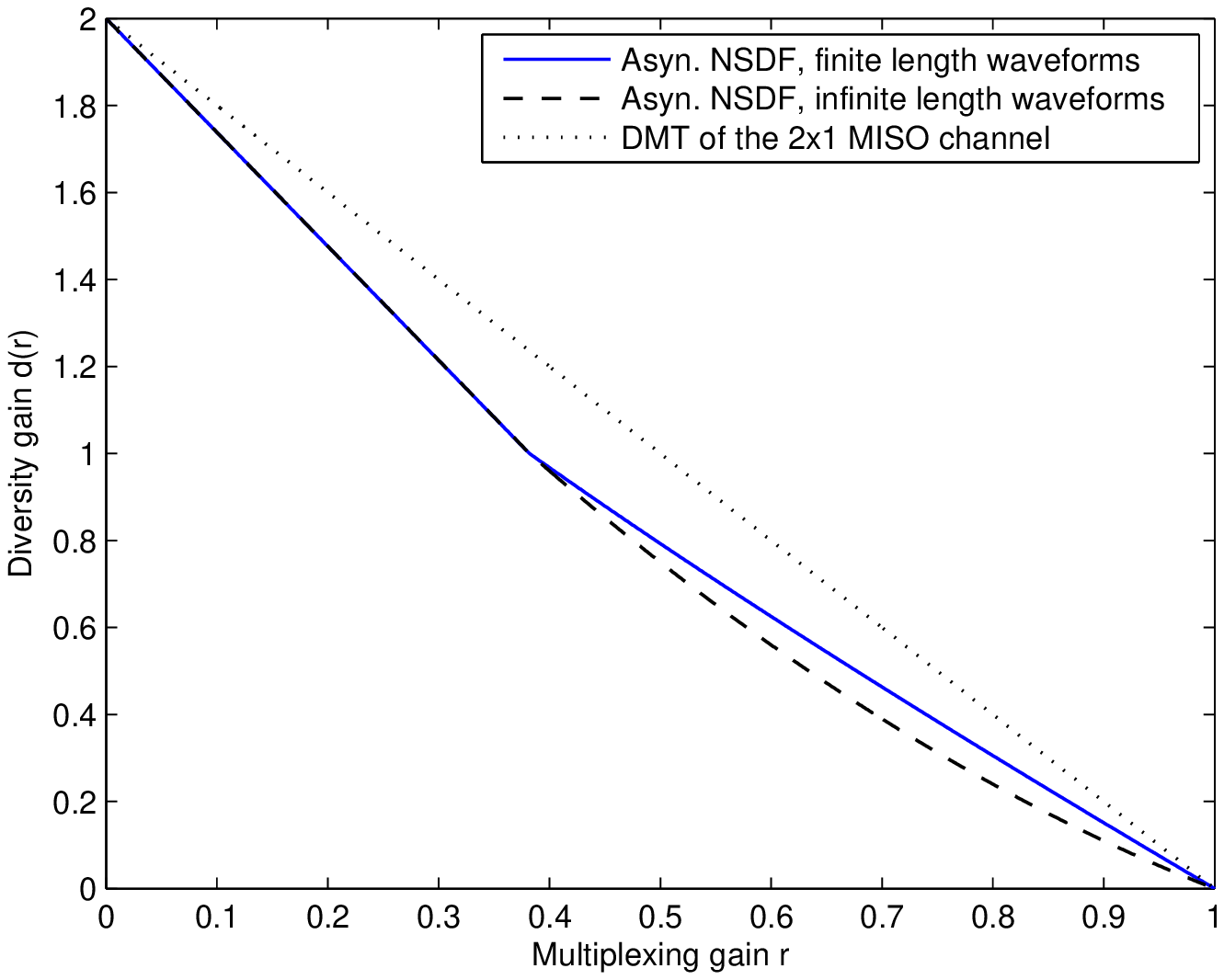}
\caption{The DMT performances of the asynchronous NSDF protocol over a single relay network for both finite and infinite length shaping waveforms and optimum values of $\kappa>1$.}\label{NSDF_OptimumK_Figure}
\end{figure}
Fig. \ref{NSDF_VariousK_Figure} illustrates the DMT performances of the NSDF protocol over the asynchronous single relay network for various values of $\kappa$ and for both scenarios of using finite length shaping waveforms (solid lines), and using infinite length shaping waveforms (dashed lines). Note that the DMT performance of the second scenario, when $u \to \infty$, is the same as that of the corresponding synchronous network. For the sake of comparison, the DMT performance of the $2 \times 1$ MISO channel is also shown (dotted line). As can be seen from this figure, for each $r$, there is a unique $\kappa$ which provides the maximum diversity gain. Fig. \ref{NSDF_OptimumK_Figure} depicts the DMT curves for the two aforementioned cases when $\kappa$ varies to maximize the diversity gain at each multiplexing gain $r$. It is observed that for $\kappa \le \frac{1+\sqrt{5}}{2}$, the DMT performances of both scenarios are the same; however, for $\kappa >\frac{1+\sqrt{5}}{2}$, the asynchronous protocol with finite length shaping waveforms provides higher diversity gain than the corresponding counterpart. Note that the extra diversity gain at high multiplexing region is at the expense of a possible bandwidth expansion at high values of SNR due to using finite length waveforms.

\section{Asynchronous OSDF Relaying Protocol}\label{Section_OSDF}
In the OSDF protocol, the source is silent in the second phased; however, the relays perform the same acts as those in the NSDF protocol. Hence, with some minor changes, the aforementioned mathematical analysis is applicable to this case. Here, asynchronism appears when at least two relays exist in the network.

\subsection{Asynchronous OSDF with Infinite Length Waveforms}
By pursuing the same procedure as that of the NSDF protocol in Section \ref{Section_NSDF_infinite_u}, the mutual information between the source and the destination when $E_m$ occurs, $0\le m\le M$, for large values of $\rho$ is given by
\begin{equation}
  I_{E_m} \doteq \frac{p}{\ell}\log(1+\rho|h_0|^2) + \frac{q}{\ell}\log\Big(1+\rho\sum_{i=1}^m|h_i|^2\Big).
\end{equation}
As can be seen, $I_{E_m}$ is the same as that of the corresponding synchronous network \cite{Elia-Vinodh-Anand-Kumar}. Hence, the DMT performances of the OSDF over both networks are the same.
\begin{theorem}
For $u \to \infty$, the DMT performance of the OSDF protocol over the underlying asynchronous cooperative relay network for a fix value of $\kappa \ge 1$ is given by\\
If $1 \le \kappa \le \frac{M+1}{M}$,
\begin{equation}\nonumber
 d^*(r) = \left\{
 \begin{array}{ll}
  (M+1)(1-\frac{\ell}{p}r), & 0 \le r \le \eta_1\\
  1-r, & \eta_1 \le r \le 1,
 \end{array}\right.
\end{equation}
else, for $\kappa \ge \frac{M+1}{M}$,
\begin{equation}\nonumber
 d^*(r) = \left\{
 \begin{array}{ll}
  (M+1)(1-\frac{M\ell}{(M+1)q}r), & 0 \le r \le \eta_2\\
  \frac{\ell}{p}(1-r), & \eta_2 \le r \le \eta_3\\
  (M+1)(1-\frac{\ell}{p} r),& \eta_3 \le r \le \eta_1\\
  1-r, & \eta_1 \le r \le 1.
 \end{array}\right.
\end{equation}
where $\eta_1 = \frac{Mp}{(M+1)\ell-p}$, $\eta_2 = \frac{q}{\ell}$, and $\eta_3 = \frac{(M+1)p-\ell}{M\ell}$. When $\kappa$ varies to maximize the diversity gain at each multiplexing gain $r$, we get
\begin{equation}\nonumber
 d^*(r) = \left\{
 \begin{array}{ll}
   (M+1)\left(1-\frac{2M+1}{M+1}r\right),& 0 \le r \le \frac{M}{2M+1}\\
   \frac{(M+1)(1-r)}{M+r+1}, & \frac{M}{2M+1} \le r \le 1.
 \end{array}\right.
\end{equation}
where the optimum $\kappa$ corresponding to each $r$ is given
by
\begin{equation}\nonumber
 \kappa = \left\{
 \begin{array}{ll}
  \frac{M+1}{M}, & 0 \le r \le \frac{M}{2M+1}\\
  \frac{1+Mr}{M(1-r)}, & \frac{M}{2M+1} \le r \le 1.
 \end{array}\right.
\end{equation}
\end{theorem}
The proof is given in \cite{Elia-Vinodh-Anand-Kumar} and is omitted here for brevity.

\subsection{Asynchronous OSDF with Finite Length Waveforms}
By pursuing the same procedure as that of the NSDF protocol in Section \ref{Section_NSDF_finite_u}, one can show that at high SNR regime the mutual information between the source and the destination when $E_m, 0 \le m \le M$, occurs is given by
\begin{align}\nonumber
  I_{E_m} &\doteq \frac{p}{\ell}\log(1+\rho|h_0|^2)+\frac{q}{\ell}\sum_{i=1}^m\log(1+\rho|h_i|^2)\\
   &\doteq \Big[\frac{p}{\ell}(1-\alpha_0)^+ + \frac{q}{\ell}\sum_{i=1}^m(1-\alpha_i)^+\Big]\log \rho.
\end{align}
Similarly, the outage probability in this case is obtained as
\begin{eqnarray}
  P_{\mathcal{O} \mid E_m} = Pr(I_{E_m} < r\log \rho) \doteq \rho^{-d_{E_m}(r)},
\end{eqnarray}
where for $\alpha_i \ge 0,\, i=0,\ldots,m,$
\begin{eqnarray}
  d_{E_m}(r) = \inf_{p(1-\alpha_0)^+ + q\sum_{i=1}^m(1-\alpha_i)^+ < \ell r} \sum_{i=0}^m\alpha_i.
\end{eqnarray}
By solving the above optimization problem, we get
\begin{lemma}\label{Lemma4}
\begin{equation}\nonumber
  d_{E_m}(r) = \left\{
  \begin{array}{ll}
    1+m-\frac{\ell}{q}r, & 0 \le r \le \frac{mq}{\ell}\\
    1+\frac{mq}{p}-\frac{\ell}{p}r, & \frac{mq}{\ell} < r \le \frac{p}{\ell}.
  \end{array}\right.
\end{equation}
Clearly, when $m \ge \kappa $, then $\frac{mq}{\ell} \ge
\frac{p}{\ell}$. In this case,
\begin{equation}\nonumber
  d_{E_m}(r) = 1 + m - \frac{\ell}{q}r, ~~~ 0 \le r \le \frac{p}{\ell}
\end{equation}
\end{lemma}

The proof is similar to that of the Lemma \ref{Lemma_dEm_NSDF_FixK} and is omitted for brevity. Here $Pr(E_m)$ is given by
\begin{equation}
  Pr(E_m) \doteq \left\{
  \begin{array}{ll}
  \rho^{-(1-\frac{\ell r}{p})(M-m)}, & 0 \le r \le \frac{p}{\ell},\\
  0, & \frac{p}{\ell} < r \le 1.
  \end{array}\right.
\end{equation}

The following theorem treats the simplest case where there are only two relays in the network.
\begin{proposition}\label{Proposition_OSDF_2Relay}
For a finite value of $u$, the DMT performance of the OSDF protocol over the underlying asynchronous cooperative network with two relays and for a fix $\kappa \ge 1$ is as follows.\\
If $1 \le \kappa < 2$,
\begin{equation}\nonumber
 d^*(r) = \left\{
\begin{array}{cc}
 3(1-\frac{\ell}{p}r), & 0 \le r \le \frac{2p}{3\ell-p}\\
 1-r, & \frac{2p}{3\ell-p} \le r \le 1,
\end{array}\right.
\end{equation}
else if $2 \le  \kappa < 3$,
\begin{equation}\nonumber
 d^*(r) = \left\{
\begin{array}{cc}
 3 - \frac{\ell^2}{pq}r, & 0 \le r \le \frac{q}{\ell}\\
 2(1-\frac{\ell}{p}r)+\frac{q}{p}, & \frac{q}{\ell} \le r \le \frac{p-q}{\ell}\\
3(1-\frac{\ell}{p}r), & \frac{p-q}{\ell} \le r \le \frac{2p}{3\ell-p}\\
1-r, & \frac{2p}{3\ell-p} \le r \le 1,
\end{array}\right.
\end{equation}
else, for $\kappa \ge 3$,
\begin{equation}\nonumber
 d^*(r) = \left\{
\begin{array}{cc}
 3 - \frac{\ell^2}{pq}r, & 0 \le r \le \frac{q}{\ell}\\
 2(1-\frac{\ell}{p}r)+\frac{q}{p}, & \frac{q}{\ell} \le r \le \frac{q(p-q)}{\ell(p-2q)}\\
 3-\frac{\ell}{q}r, & \frac{q(p-q)}{\ell(p-2q)} \le r \le \frac{2q}{\ell}\\
 1-\frac{\ell}{p}r+\frac{2q}{p}, & \frac{2\ell}{q} \le r \le \frac{p-q}{\ell}\\
 3(1-\frac{\ell}{p}r), & \frac{p-q}{\ell} \le r \le \frac{2p}{3\ell-p}\\
 1-r, & \frac{2p}{3\ell-p} \le r \le 1.
\end{array}\right.
\end{equation}
When $\kappa$ varies to maximize the diversity gain at each multiplexing gain $r$, we have
\begin{equation}\nonumber
d^*(r) = \left\{
\begin{array}{cc}
  3\left(1-\frac{3}{2}r\right), & 0 \le r \le \frac{1}{3}\\
  \frac{3(1-r)}{1+r}, & \frac{1}{3} \le r \le 1.
\end{array}\right.
\end{equation}
The optimum value of $\kappa$ corresponding to each multiplexing gain $r$ is given by
\begin{equation}\nonumber
 \kappa = \left\{
\begin{array}{cc}
 2, & 0 \le r \le \frac{1}{3} \\
 \frac{1+r}{1-r} , & \frac{1}{3} \le r \le 1.
\end{array}\right.
\end{equation}
\end{proposition}

The proof is given in Appendix \ref{Proof_Proposition_OSDF_2Relay}. To extend the above results to the general case, let $d^*_{M}(r)$ be the DMT performance of the OSDF protocol over an $M$ relay asynchronous cooperative network when the cooperation is not stopped throughout the range of the multiplexing gain. The following theorem concludes the results.
\begin{theorem}\label{Theorem_OSDF_MRelay}
For a finite value of $u$, the DMT performance of the OSDF protocol over an asynchronous two-hop cooperative network with $M$ relays for a fix $\kappa$ is given by\\
If $\kappa \le M+1$,
\begin{equation}\nonumber
d_M^*(r) = \left\{
\begin{array}{ll}
(1-\frac{\ell}{p}r)+d^*_{M-1}(r), & 0 \le r \le \frac{p}{\ell}\\
0, & \frac{p}{\ell} \le r \le 1,
\end{array}\right.
\end{equation}
else, for $\kappa > M+1$,
\begin{equation} \nonumber
d_M^*(r) = \left\{
\begin{array}{ll}
(1-\frac{\ell}{p}r)+d^*_{M-1}(r), & 0 \le r \le \eta_1\\
1+M-\frac{\ell}{q}r, & \eta_1 \le r \le \eta_2\\
1+\frac{Mq}{p}-\frac{\ell}{p}r, & \eta_2 \le r \le \eta_3\\
(M+1)(1-\frac{\ell}{p}r), & \eta_3 \le r \le \eta_4\\
0, & \eta_4 \le r \le 1,
\end{array}\right.
\end{equation}
where $\eta_1 = \frac{(M-1)(p-q)q}{\ell(p-2q)}$, $\eta_2 = \frac{Mq}{\ell}$, $\eta_3 = \frac{p-q}{\ell}$, and $\eta_4 = \frac{p}{\ell}$. The resulted DMT for each region of $\kappa$ is compared to $(1-r)$ to decide when to stop the cooperation.

When $\kappa$ varies to maximize the diversity gain at each multiplexing gain $r$, we obtain
\begin{equation}\nonumber
d^*(r) = \left\{
\begin{array}{ll}
  (M+1)\left(1-\frac{3}{2}r\right), & 0 \le r \le \frac{1}{3}\\
  (M+1)\frac{1-r}{1+r}, & \frac{1}{3} \le r \le 1.\\
\end{array}\right.
\end{equation}
The optimum $\kappa$ for each $r$ is given by
\begin{equation}\nonumber
 \kappa = \left\{
\begin{array}{ll}
 2, & 0 \le r \le \frac{1}{3} \\
 \frac{1+r}{1-r} , & \frac{1}{3} \le r \le 1.
\end{array}\right.
\end{equation}
\end{theorem}

The proof is given in Appendix \ref{Proof_Theorem_OSDF_MRelay}.
\begin{figure}
\centering
\includegraphics[width=9cm, height = 7cm]{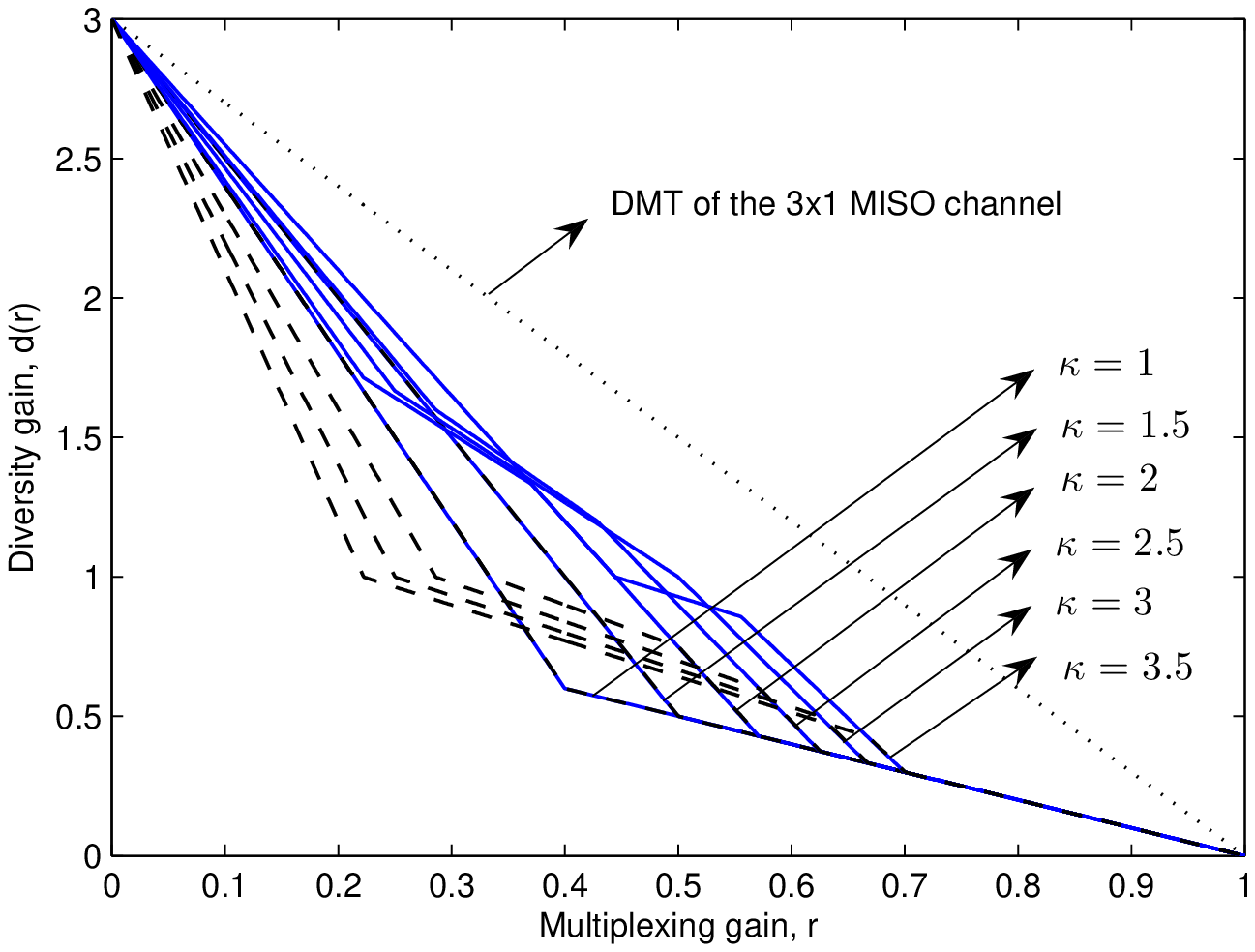}
\caption{The DMT performances of the asynchronous OSDF protocol over a two relay network for both finite and infinite length shaping waveforms and for various values of $\kappa > 1$.}\label{OSDF_FixK_Figure}
\centering
\includegraphics[width=9cm, height = 7cm]{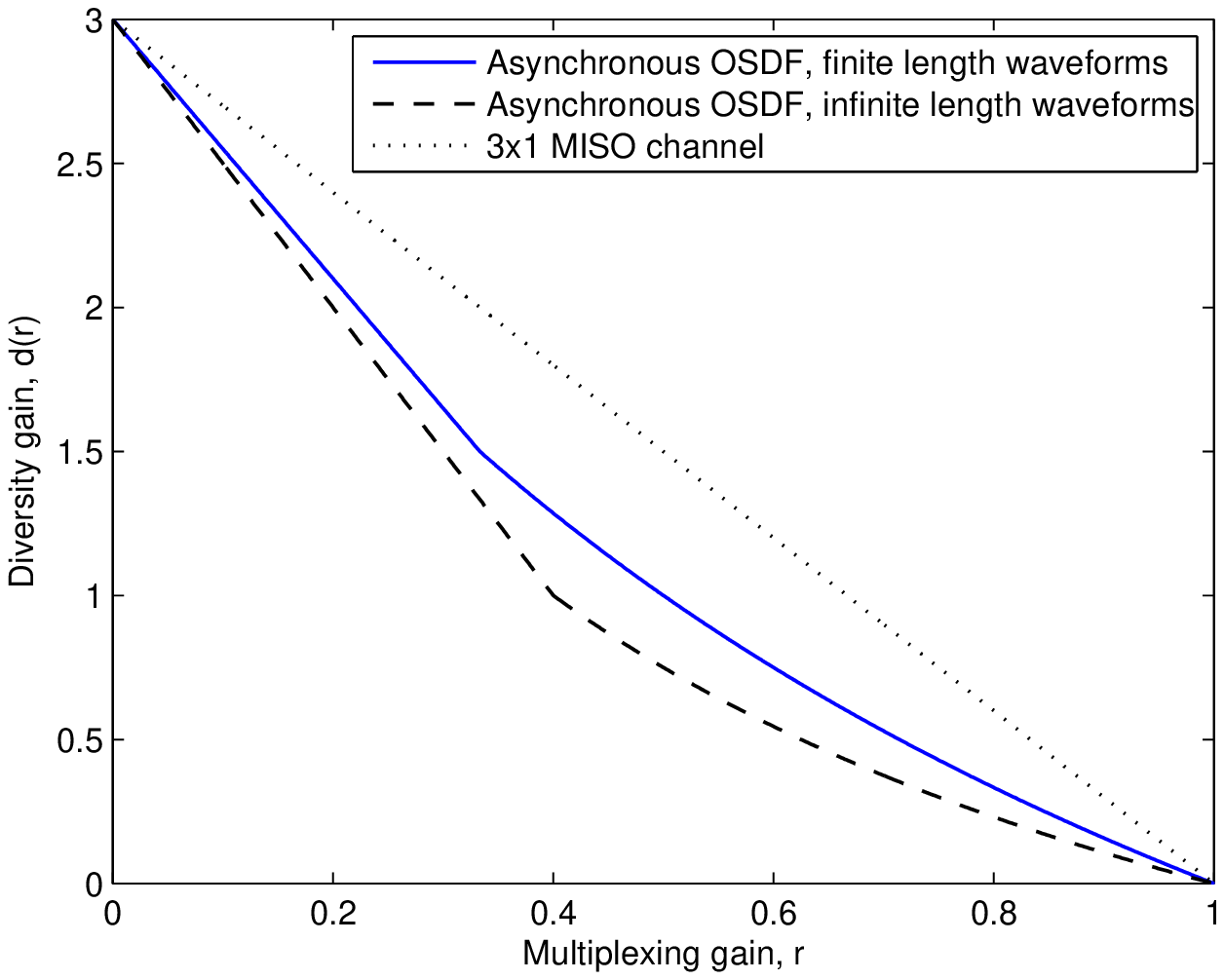}
\caption{The DMT performances of the asynchronous OSDF protocol over a two relay network for both finite and infinite length shaping waveforms and optimum values of $\kappa>1$.}\label{OSDF_VariableK_Figure}
\end{figure}
Fig. \ref{OSDF_FixK_Figure} illustrates the DMT performances of the OSDF protocol over the asynchronous two relay network for various values of $\kappa$ and for both scenarios of using finite length shaping waveforms (solid lines), and using infinite length shaping waveforms (dashed lines). Note that the DMT performance in the second scenario, when $u \to \infty$, is the same as that of the corresponding synchronous network. For comparison, the DMT of the $3 \times 1$ MISO channel is also shown (dotted line). As can be seen from this figure, for each $r$, there is a unique $\kappa$ which provides the maximum diversity gain. Fig. \ref{OSDF_VariableK_Figure} depicts the DMT curves for the two aforementioned cases when $\kappa$ varies to maximize the diversity gain at each multiplexing gain $r$. It is observed that the asynchronous protocol with finite length shaping waveforms provides higher diversity gain than the corresponding counterpart throughout the range of the multiplexing gain. It is worth nothing that the extra diversity gain is at the expense of a possible bandwidth expansion at high values of SNR.

\section{Asynchronous NAF Relaying Protocol}\label{Section_NAF}
In the second phase of the AF type protocols, the relays perform linear processing (not decoding) on the received signals and retransmit them to the destination. If $\underline{y}_{r_i}$ is the received signal vector at the $i$-th relay in the first phase, the transmitted vector $\underline{x}_i$ from this node is modeled by
\begin{equation}
 \underline{x}_i = \mathbf{A}_i\underline{y}_{r_i},
\end{equation}
where $\mathbf{A}_i$ is a $q \times p$ matrix of rank $q \le p$. In the NAF protocol, the source sends a new codeword of length $q$ to the destination in the second phase.

\subsection{Asynchronous NAF with Infinite Length Waveforms}\label{Section_NAF_infinite_u}
If $\underline{x}'_0$ is the source's transmitted codeword in the first phase, the received signal vectors at the $i$-th relay and the destination are given by
\begin{align}
 \underline{y}_{r_i} &= g_i\underline{x}'_0+\underline{z}_{r_i},\\
 \underline{y}'_{d} &= h_0\underline{x}'_0+\underline{z}'_{d},
\end{align}
where all vectors are of length $p$. $\underline{z}_{r_i}$ and $\underline{z}'_{d}$ are the additive white Gaussian noise vectors at the $i$-th relay and at the destination in the first phase.

The received signals at the relays are linearly processed and retransmitted to the destination. At the destination, the received signal vector in the second phase according to \eqref{Eqn-yd-infiniteu} is given by
\begin{equation}
  \underline{y}_{d} = \sum_{j=0}^M h_j\mathbf{\Gamma}_{j}\underline{x}_j + \underline{z}_d,
\end{equation}
where $\mathbf{\Gamma}_j$ is given in \eqref{Eqn_Gamma_j}. By replacing $\underline{x}_j = \mathbf{A}_j(g_j\underline{x}'_0+\underline{z}_{r_j})$ for $j=1,2,\ldots,M$, we obtain
\begin{align} \nonumber
  \underline{y}_{d} =&  h_0\underline{x}_0 + \Big(\sum_{j=1}^M h_j g_j \mathbf{\Gamma}_j\mathbf{A}_j\Big) \underline{x}'_0 + \sum_{j=1}^M h_j \mathbf{\Gamma}_j\mathbf{A}_j\underline{z}_{r_j} + \underline{z}_d,
\end{align}
The system model for both phases is given by
\begin{equation}
 \underline{y} = \mathbf{H}\underline{x}+\underline{z},
\end{equation}
where
\begin{align*}
 \underline{y} &= \Big[\big(\underline{y}'_{d}\big)^T, \underline{y}_{d}^T\Big]^T,\\
 \underline{z} &= \Big[\big(\underline{z}'_{d}\big)^T, \underline{c}^T + \underline{z}_{d}^T\Big]^T,\\
 \underline{x} &= \Big[\big(\underline{x}'_0\big)^T,\underline{x}_0^T\Big]^T,\\
 \nonumber \mathbf{H} &= \left[
 \begin{array}{cc}
   h_0\mathbf{I}_{p} & \mathbf{0}_{p \times q}\\
   \mathbf{G} & h_0\mathbf{I}_q
 \end{array}\right],
\end{align*}
$\underline{c} = \sum_{j=1}^M h_j \mathbf{\Gamma}_j\mathbf{A}_j\underline{z}_{r_j}$, and $\mathbf{G} = \sum_{j=1}^M h_j g_j \mathbf{\Gamma}_j\mathbf{A}_j$. The covariance matrix of the noise vector $\underline{z}$ is given by
\begin{equation}
  \mathbf{\Phi} = \sigma^2_d\left[
  \begin{array}{cc}
    \mathbf{I}_p & \mathbf{0}_{p \times q}\\
    \mathbf{0}_{q \times p} & \mathbf{C}
  \end{array}\right],
\end{equation}
where $\mathbf{C} = \mathbf{I}_q+\frac{\sigma^2_r}{\sigma^2_d}\sum_{j=1}^M|h_j|^2\mathbf{\Gamma}_j\mathbf{A}_j
\mathbf{A}_j^\dagger\mathbf{\Gamma}_j^\dagger$. If the codebooks are Gaussian, the mutual information between the source and the destination is given by
\begin{align}\nonumber
  I(\underline{x};\underline{y}) &= \log \det (\mathbf{I}_\ell+\mathbf{H\Sigma}_x\mathbf{H}^\dagger\mathbf{\Phi}^{-1})\\
  & = (1+\rho|h_0|^2)^p\det(\mathbf{C}^{-1}) \det\Big[\mathbf{C}+\rho|h_0|^2\mathbf{I}_q+\frac{\rho}{1+\rho|h_0|^2}
  \mathbf{GG}^\dagger\Big],
\end{align}
where $\mathbf{\Sigma}_x$ is the autocorrelation matrix of the input vector $\underline{x}$ which is assumed to be equal to $\mathcal{E}\mathbf{I}_\ell$, where $\mathcal{E}$ is the average transmitted energy per symbol. It is shown in \cite{Elia-Vinodh-Anand-Kumar} that
\begin{align*}
  \mathbf{GG}^\dagger &\preceq M\sum_{j=1}^M|h_jg_j|^2
  \mathbf{\Gamma}_j\mathbf{A}_j\mathbf{\Gamma}_j^\dagger\mathbf{A}_j^\dagger.
\end{align*}
Moreover, since $\mathbf{C} \succeq \mathbf{I}_q$, we have $\det{\mathbf{C}^{-1}} \le 1$. Let $\mathcal{A} \doteq \mathbf{I}_\ell+\mathbf{H\Sigma}_x\mathbf{H}^\dagger\mathbf{\Phi}^{-1}$. By proceeding in the footsteps of \cite{Elia-Vinodh-Anand-Kumar}, we get
\begin{equation}
  \det \mathcal{A} \,\dot\le \, (1+\rho|h_0|^2)^p\left(1+\rho|h_0|^2+\sum_{j=1}^M |h_j|^2+\frac{\rho|h_jg_j|^2}{1+\rho|h_0|^2}\right)^q
\end{equation}
It is shown in \cite{Elia-Vinodh-Anand-Kumar} that by proper choice of the $\mathbf{A}_j$ matrices, this bound is achievable and is in fact tight.
Define $\alpha_j \doteq -\frac{\log|h_j|^2}{\log\rho}$, and $\beta_j \doteq \frac{\log |h_jg_j|^2}{\log\rho}$. Let $\beta = \min_{j\ge 1} \beta_j$ and $\alpha = \min_{j\ge 1} \alpha_j$. We get
\begin{align}
  I(\underline{x};\underline{y})\, \doteq & \, \big[(p-q)(1-\alpha_0)^+ + q \max\{-\alpha, 2(1-\alpha_0), (1-\alpha-\alpha_0), (1-\beta)\}^+ \big]\log \rho.
\end{align}
By proceeding in the footsteps of \cite{Zheng-Tse}, the outage probability at high values of SNR is given by
\begin{align}
  P_{\mathcal{O}} = Pr(I(\underline{x};\underline{y}) < \ell r\log \rho)
  \doteq \rho^{-d^*(r)},
\end{align}
where
\begin{equation}
  d^*(r) = \inf_{\mathcal{R}} \alpha_0+M\alpha+M\beta,
\end{equation}
and $\mathcal{R} = \{(p-q)(1-\alpha_0)^+ + q \max\{-\alpha, 2(1-\alpha_0), (1-\alpha-\alpha_0),(1-\beta)\}^+ < \ell r,\, \alpha_0, \alpha,\beta \ge 0\}$.
Clearly, it is sufficient to consider $0 \le \alpha_0, \beta \le 1$. Moreover, since $\alpha \ge 0$, we simply set it to zero to get
\begin{equation}
  d^*(r) = \inf_{(p-q)\alpha_0+q\min\{2\alpha_0-1,\beta\} > p-\ell r} \alpha_0+M\beta.
\end{equation}
By solving the above optimization problem, we obtain
\begin{theorem}\label{Theorem_NAF_infinite_p}
For $u \to \infty$, the DMT performance of the NAF protocol over the underlying asynchronous cooperative network for a fix value of $\kappa \ge 1$ is as follows.\\
If $1 \le \kappa \le \frac{M+1}{M}$.
\begin{equation*}
  d^*(r) = M(1-2r)^++(1-r), ~~ 0 \le r \le 1.
\end{equation*}
Else, for $\kappa \ge \frac{M+1}{M}$
  \begin{equation*}
    d^*(r) = \left\{
    \begin{array}{ll}
      \big(1- \frac{M(p-q)}{q}r\big) + M(1-2r), & 0\le r \le \frac{q}{\ell}\\
      (1- r) + \frac{q}{p-q}(1-2r), & \frac{q}{\ell} \le r \le \frac{1}{2}\\
      1- r, & \frac{1}{2} \le r \le 1.
    \end{array}\right.
  \end{equation*}
\end{theorem}
The best DMT is achieved when $1\le \kappa \le \frac{M+1}{M}$.

The proof is given in Appendix \ref{Apx_proof_Theorem_NAF_infinite_p}.
It is seen that the best DMT performance of the NAF protocol over the underlying asynchronous network is the same as the DMT of this protocol over the corresponding synchronous network. Hence, the asynchronism does not diminish the DMT performance of the underlying network.

\subsection{Asynchronous NAF with Finite Length Waveforms}\label{Section_NAF_finite_u}
If $\underline{x}'_0$ is the source's transmitted codeword in the first phase, the received signal vectors at the $i$-th relay and the destination are given by
\begin{align}
 \underline{y}_{r_i} &= g_i\mathbf{\Gamma}'_{0,0}\underline{x}'_0+\underline{z}_{r_i},\\
 \underline{y}'_{d,0} &= h_0\mathbf{\Gamma}'_{0,0}\underline{x}'_0+\underline{z}'_{d,0},
\end{align}
where all vectors are of length $p$. $\mathbf{\Gamma}'_{0,0}$ of size $p \times p$ represents the effect of the ISI among the source's transmitted symbols at phase one. $\underline{z}_{r_i}$ and $\underline{z}'_{d,0}$ are the additive Gaussian noise vectors at the $i$-th relay and at the destination in the first phase with the covariance matrices $\sigma^2_r\mathbf{\Gamma}'_{0,0}$, $\sigma^2_d\mathbf{\Gamma}'_{0,0}$, respectively.

The received signals at the relays are linearly processed and retransmitted to the destination. The output matched filters are indexed from $0$ to $M$ where the $0$-th filter is matched on the link between the source and the destination. The received signal vector at the output of the $i$-th matched filter in the second phase according to \eqref{Eqn_y_di_finite_u} is given by
\begin{equation}
  \underline{y}_{d,i} = \sum_{j=0}^M h_j\mathbf{\Gamma}_{i,j}\underline{x}_j + \underline{z}_i,~~~i=0,1,\ldots,M.
\end{equation}
By replacing $\underline{x}_j = \mathbf{A}_j(g_j\mathbf{\Gamma}'_{0,0}\underline{x}'_0+\underline{z}_{r_j})$ for $j=1,2,\ldots,M$, we obtain
\begin{align}
  \underline{y}_{d,i} =&  h_0\mathbf{\Gamma}_{0,0}\underline{x}_0 + \Big(\sum_{j=1}^M h_j g_j \mathbf{\Gamma}_{i,j}\mathbf{A}_j \mathbf{\Gamma}'_{0,0}\Big) \underline{x}'_0 + \sum_{j=1}^M h_j g_j \mathbf{\Gamma}_{i,j}\mathbf{A}_j\underline{z}_{r_j} +  \underline{z}_i.
\end{align}

The system model for both phases is given by
\begin{equation}
 \underline{y} = \mathbf{H}\underline{x}+\underline{z},
\end{equation}
where
\begin{align*}
 \underline{y} &= \Big[(\underline{y}'_{d,0})^T, \underline{y}_{d,0}^T,\underline{y}_{d,1}^T, \ldots, \underline{y}_{d,M}^T\Big]^T,\\
 \underline{z} &= \Big[(\underline{z}'_{d,0})^T, \underline{c}_0^T + \underline{z}_{d,0}^T, \underline{c}_1^T + \underline{z}_{d,1}^T,\ldots,\underline{c}_M^T + \underline{z}_{d,M}^T\Big]^T,\\
 \underline{x} &= \Big[\big(\underline{x}'_0\big)^T,\underline{x}_0^T\Big]^T,\\
 \nonumber \mathbf{H} &= \left[
 \begin{array}{cc}
   h_0\mathbf{\Gamma}'_{0,0} & \mathbf{0}_{p \times q}\\
   \bf{G} & h_0\mathbf{\Gamma}
 \end{array}\right].
\end{align*}
$\mathbf{G} = [\mathbf{G}_0^T,\mathbf{G}_1^T,\ldots,\mathbf{G}_M^T]^T$ and $\mathbf{\Gamma} = [\mathbf{\Gamma}_{0,0}^T, \mathbf{\Gamma}_{1,0}^T,\ldots, \mathbf{\Gamma}_{M,0}^T]^T$, where for $i=0,1,\ldots,M$,
\begin{align*}\nonumber
\mathbf{G}_i &= \sum_{j=1}^Mh_jg_j\mathbf{\Gamma}_{i,j}\mathbf{A}_j\mathbf{\Gamma}'_{0,0},\\
\underline{c}_i & = \sum_{j=1}^M h_j\mathbf{\Gamma}_{i,j}\mathbf{A}_j\underline{z}_{r_{j}}.
\end{align*}

The covariance matrix of the noise is calculated as
\begin{equation}
 \mathbf{\Phi} = \sigma_d^2 \left[
 \begin{array}{cc}
  \mathbf{\Gamma}'_{0,0} & \mathbf{0}_{p \times (M+1)q}\\
  \mathbf{0}_{(M+1)q \times p} & \mathbf{C}
 \end{array}\right],
\end{equation}
where $\mathbf{C} = \big[\mathbf{C}_{i,j}\big], \, i,j = 0,1,\ldots,M$, and
\begin{equation} \nonumber
 \mathbf{C}_{i,j} = \mathbf{\Gamma}_{i,j} + \frac{\sigma_r^2}{\sigma_d^2}
  \sum_{k=1}^M |h_k|^2\mathbf{\Gamma}_{i,k}\mathbf{A}_{k}\mathbf{\Gamma}'_{0,0}\mathbf{A}_{k}^\dagger\mathbf{\Gamma}_{j,k}^\dagger.
\end{equation}

Define
\begin{align}
 \mathbf{\Xi} &\triangleq \left[
\begin{array}{ccccc}
 \mathbf{\Gamma}_{0,0} & \mathbf{\Gamma}_{0,1} & \ldots & \mathbf{\Gamma}_{0,M}\\
 \mathbf{\Gamma}_{1,0} & \mathbf{\Gamma}_{1,1} &\ldots & \mathbf{\Gamma}_{1,M}\\
 \vdots & \vdots &  & \vdots\\
 \mathbf{\Gamma}_{M,0} & \mathbf{\Gamma}_{M,1} & \ldots & \mathbf{\Gamma}_{M,M}
\end{array}\right],\\
\mathbf{\Sigma} &\triangleq
 \big[h_1g_1\mathbf{A}_1^T, h_2g_2\mathbf{A}_2^T, \ldots, h_Mg_M\mathbf{A}_M^T \big]^T.
\end{align}
One can check that
\begin{align}
    \mathbf{G} &= \mathbf{\Xi}[\mathbf{0}_{p\times q}, (\mathbf{\Sigma}\mathbf{\Gamma}'_{0,0})^\dagger]^\dagger,\\
    \mathbf{\Gamma\Gamma}^\dagger &= \mathbf{\Xi}\texttt{diag}\{\mathbf{I}_q,
    \mathbf{0}_{Mq\times Mq}\}\mathbf{\Xi}\\
    \mathbf{C} & = (\mathbf{\Xi}\texttt{diag}\{\mathbf{0},\hat{\mathbf{A}}_1,\ldots, \hat{\mathbf{A}}_M\} + \mathbf{I}_{(M+1)q})\mathbf{\Xi},
\end{align}
where $\hat{\mathbf{A}}_i = \frac{\sigma_r^2}{\sigma_d^2}|h_i|^2\mathbf{A}_i\mathbf{\Gamma}'_{0,0}\mathbf{A}_i^\dagger$.
Hence, $\mathbf{C}^{-1}$ exists if and only if $\mathbf{\Xi}^{-1}$ exists. According to Proposition \ref{Prop_det_Xi_k}, if the shaping waveforms $\psi_i(t), i=0,\ldots, M$, are designed properly, $\mathbf{\Xi}$ is a positive definite matrix and $\mathbf{\Xi}^{-1}$ exists. Assuming $\psi_0(t)$ is a well designed waveform with non-zero spectrum over its bandwidth,  $\mathbf{\Gamma}'_{0,0}$ is also a full rank matrix with bounded positive real eigenvalues (see \cite{Robert-Gray}). Therefore,
$\mathbf{\Phi}^{-1}$ is given by
\begin{equation}
 \mathbf{\Phi}^{-1} = \frac{1}{\sigma_d^2}~\texttt{diag}\{(\mathbf{\Gamma}'_{0,0})^{-1}, \mathbf{C}^{-1}\}.
\end{equation}

Let $\mathcal{A} \triangleq \mathbf{I}_{p+(M+1)q}+\mathbf{H\Sigma}_x\mathbf{H}^\dagger\mathbf{\Phi}^{-1}$. The mutual information between the source and the destination is given by
\begin{equation}
 I(\underline{x};\underline{y}) = \log \det\mathcal{A}.
\end{equation}
$\mathcal{A}$ is given by
\begin{equation*}
\mathcal{A} = \left[
\begin{array}{cc}
 \mathbf{I}_p+\rho|h_0|^2\mathbf{\Gamma}'_{0,0} & \rho h_0\mathbf{\Gamma}'_{0,0}\mathbf{G}^\dagger\mathbf{C}^{-1}\\
 \rho h_0^*\mathbf{G} & \mathbf{I}_{(M+1)q} + \rho(\mathbf{G}\mathbf{G}^\dagger
 + |h_0|^2\mathbf{\Gamma\Gamma}^\dagger)\mathbf{C}^{-1}
\end{array}\right].
\end{equation*}
The determinant of $\mathcal{A}$ is given by
\begin{align}\nonumber
 \det\mathcal{A}  = & \det\big(\mathbf{I}_p+\rho|h_0|^2\mathbf{\Gamma}'_{0,0}\big)
 \det\Big(\mathbf{I}_{(M+1)q}+ \rho|h_0|^2\mathbf{\Gamma\Gamma}^\dagger\mathbf{C}^{-1} + \rho\mathbf{G}\mathbf{B}\mathbf{G}^\dagger\mathbf{C}^{-1}\Big)
\end{align}
where $\mathbf{B} = \big(\mathbf{I}_p + \rho|h_0|^2\mathbf{\Gamma}'_{0,0}\big)^{-1}$.
It can be checked that
\begin{align*}
\mathbf{G}\mathbf{B}\mathbf{G}^\dagger\mathbf{C}^{-1} &= \mathbf{\Xi}~\texttt{diag}\{\mathbf{0}_{q\times q},\mathbf{\Sigma}\mathbf{\Gamma}'_{0,0}\mathbf{B}
(\mathbf{\Gamma}'_{0,0})^\dagger\Sigma^\dagger\}\mathbf{\Psi},\\
\mathbf{\Gamma\Gamma}^\dagger\mathbf{C}^{-1} &=
\mathbf{\Xi}\texttt{diag}\{\mathbf{I}_q, \mathbf{0}_{Mq \times Mq}\}\mathbf{\Psi},
\end{align*}
where $\mathbf{\Psi} = \left(\mathbf{\Xi}\texttt{diag}\{0,\hat{\mathbf{A}}_1,\ldots,\hat{\mathbf{A}}_M\}+\mathbf{I}_{(M+1)q}
\right)^{-1}$.
Hence,
\begin{align}\nonumber
 \det\mathcal{A}  = & \det\big(\mathbf{I}_p+\rho|h_0|^2\mathbf{\Gamma}'_{0,0}\big)
 \det\Big(\mathbf{I}_{(M+1)q}+ \rho\mathbf{\Xi}\texttt{diag}\{|h_0|^2\mathbf{I}_q,\mathbf{\Sigma}
\mathbf{\Gamma}'_{0,0}\mathbf{B} (\mathbf{\Gamma}'_{0,0})^\dagger\Sigma^\dagger\}\mathbf{\Psi}\Big).
\end{align}
Since $\mathbf{\Xi}$, $\mathbf{\Psi}$, and $\mathbf{\Gamma}'_{0,0}$ are positive definite matrices with all bounded eigenvalues, they do not affect the mutual information when $\rho \to \infty$. We obtain,
\begin{align}\nonumber
 \det\mathcal{A} \doteq &(1+\rho|h_0|^2)^{p+q}\det(\mathbf{I}_q+\rho\mathbf{\Sigma}^\dagger
\mathbf{\Sigma}\mathbf{B})\\
\doteq & (1+\rho|h_0|^2)^{p}\Big(1+\rho|h_0|^2+\rho\sum_{j=1}^M |h_jg_j|^2\Big)^q
\end{align}

Let $\alpha_0 \triangleq -\frac{\log|h_0|^2}{\log \rho}$, $\beta_i \triangleq -\frac{\log|h_ig_i|^2}{\log \rho}$, and $\beta \triangleq \min_{i\ge 1} \beta_i$. We obtain,
\begin{equation}
 I(\underline{x};\underline{y}) \doteq \left[p(1-\alpha_0)^+ + q\max\{1-\alpha_0, 1-\beta\}^+\right]\log\rho.
\end{equation}
By proceeding in the footsteps of \cite{Zheng-Tse}, the outage probability is given by
\begin{equation}\nonumber
  P_{\mathcal{O}}(R) = Pr(I(\underline{x};\underline{y}) < \ell R) \doteq \rho^{-d^*(r)},
\end{equation}
where for $\alpha_i, \beta_i \ge 0,\, \forall i \in \{0,1,\ldots,M\}$,
\begin{equation}
  d^*(r) = \inf_{p(1-\alpha_0)^+ + q\max\{1-\alpha_0,1-\beta\}^+< \ell r} \alpha_0+M\beta.
\end{equation}
Clearly, $\inf (\alpha_0+M\beta)$ occurs when $0 \le \alpha_0, \beta \le 1$. Hence,
\begin{equation}
  d^*(r) = \inf_{p\alpha_0 + q \min\{\alpha_0,\beta\} > \ell(1-r)} \alpha_0+M\beta.
\end{equation}
By solving the above optimization problem, we get
\begin{theorem}\label{Theorem_MainResult_M_Relay_NAF}
For a finite value of $u$, the DMT performance of the NAF protocol over the underlying asynchronous cooperative network with $M$ relays for a fix $\kappa \ge 1$ is given by
\begin{equation}\nonumber
 d^*(r) = \left\{
\begin{array}{cc}
 (M+1)(1-\frac{M\ell}{(M+1)q}r), & 0 \le r \le \frac{q}{\ell}\\
 1+\frac{q}{p}-\frac{\ell}{p}r, & \frac{q}{\ell} \le r \le 1.
\end{array}\right.
\end{equation}
The best DMT is achieved when $\kappa = 1$. In this case, for large length codewords
\begin{equation}\nonumber
  d^*(r) = \left\{
  \begin{array}{cc}
    (M+1)(1-\frac{2M}{(M+1)}r), & 0 \le r \le \frac{1}{2}\\
    2(1-r), & \frac{1}{2} \le r \le 1.
  \end{array}\right.
\end{equation}
\end{theorem}

The proof is given in Appendix \ref{Apx_Proof_Theorem_NAF_finite_u}.
Fig. \ref{NAF_OptimumK_Figure} depicts the DMT curves of the NAF protocol over a single relay asynchronous cooperative network for both cases of using finite length shaping waveforms (solid line), and using infinite length shaping waveforms (dashed line) for the optimum value of $\kappa=1$. Note that the DMT performance in the latter case is the same as that of the corresponding synchronous network. As can be seen, the asynchronous network with finite length shaping waveforms provides the same DMT performance as that of a
$2 \times 1$ MISO channel. Obviously, the extra gain is achieved at the expense of a possible bandwidth expansion at high values of SNR.
\begin{figure}
\centering
\includegraphics[width=9 cm , height = 7cm]{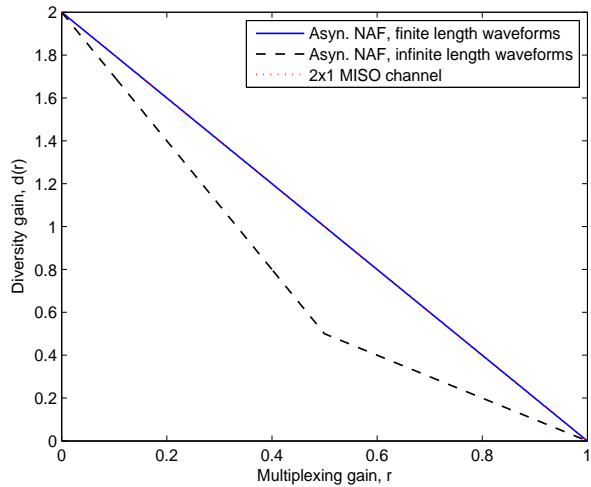}
\caption{The DMT performances of the asynchronous NAF protocol for both finite and infinite length shaping waveforms and optimum values of $\kappa=1$.}\label{NAF_OptimumK_Figure}
\end{figure}

\section{Asynchronous OAF Relaying Protocol}\label{Section_OAF}
In the OAF protocol, the source becomes silent in the second phase; however, the relays perform the same acts as those of the NAF protocol. Hence, with some minor changes, the mathematical analysis presented in
Section \ref{Section_NAF} can be used here. Since the protocol is orthogonal, asynchronism appears when at least two relays are in the network.

\subsection{Asynchronous OAF with Infinite Length Waveforms}\label{Subsec_OAF_infinite_u}
By pursuing the same procedure as that of the NAF protocol presented in Section \ref{Section_NAF_infinite_u}, the mutual information between the source and the destination for large values of SNR is given by
\begin{equation}
  I(\underline{x};\underline{y}) \dot\le \log (1+\rho|h_0|^2)^p\left(1+\sum_{j=1}^M|h_j|^2+\frac{\rho|h_jg_j|^2}{1+\rho|h_0|^2}\right)^q.
\end{equation}
It is shown in \cite{Elia-Vinodh-Anand-Kumar} that this upper bound is achievable and in fact is tight.  Define $\alpha_j \doteq -\frac{\log|h_j|^2}{\log\rho}$, and $\beta_j \doteq \frac{\log |h_jg_j|^2}{\log\rho}$. Let $\beta = \min_{j\ge 1} \beta_j$ and $\alpha = \min_{j\ge 1} \alpha_j$. We get
\begin{align}
  I(\underline{x};\underline{y})\, \doteq & \, \big[(p-q)(1-\alpha_0)^+ + q \max\{-\alpha,(1-\alpha_0), (1-\alpha-\alpha_0), (1-\beta)\}^+ \big]\log \rho.
\end{align}
By proceeding in the footsteps of \cite{Zheng-Tse}, the outage probability at high values of SNR is given by
\begin{align}
  P_{\mathcal{O}} = Pr(I(\underline{x};\underline{y}) < \ell r\log \rho)
  \doteq \rho^{-d^*(r)},
\end{align}
where by considering $0 \le \alpha_0, \beta \le 1$ and $\alpha = 0$,
\begin{equation}\label{Eqn_d*(r)_OAF_Infinite_p}
  d^*(r) = \inf_{(p-q)\alpha_0+q\min\{\alpha_0,\beta\}\ge p-\ell r} \alpha_0+M\beta.
\end{equation}
One can see that $I(\underline{x};\underline{y})$ of the underlying asynchronous network under OAF protocol is the same as that of the corresponding synchronous network under the same protocol. Hence, both networks provide the same DMT performances.
\begin{theorem}\label{Theorem_OAF_infinite_p}
For $u \to \infty$, the DMT performance of the OAF protocol over the underlying asynchronous network for a fix $\kappa \ge 1$ is given as follows.\\
If $\kappa \le \frac{M+1}{M}$
\begin{equation}\nonumber
 d^*(r) = \left\{
\begin{array}{ll}
 (M+1)(1-\frac{\ell}{p}r), & 0 \le r \le \frac{p}{\ell}\\
 0, & \frac{p}{\ell} \le r \le 1,
\end{array}\right.
\end{equation}
else if $\kappa \ge \frac{M+1}{M}$
\begin{equation}\nonumber
 d^*(r) = \left\{
\begin{array}{ll}
 (M+1)(1-\frac{M\ell}{(M+1)q}r), & 0 \le r \le \frac{q}{\ell}\\
 \frac{p}{p-q}(1-\frac{\ell}{p}r), & \frac{q}{\ell} \le r \le \frac{p}{\ell} \\
0, & \frac{p}{\ell} \le r \le 1.
\end{array}\right.
\end{equation}
The best DMT for $0 \le r \le \frac{1}{2}$ is achieved when $\kappa = \frac{M+1}{M}$. For $\frac{1}{2}\le r \le 1$, the best DMT is achieved when the source transmits alone. Hence,
\begin{align*}
  d^*(r) &= \left\{
  \begin{array}{ll}
    (M+1)(1-\frac{2M+1}{M+1}r), & 0 \le r \le \frac{1}{2}\\
    1-r, & \frac{1}{2} \le r \le 1.
  \end{array}\right.\\
  & = M(1-2r)^+ + (1-r)^+.
\end{align*}
\end{theorem}

The proof is given in \cite{Elia-Vinodh-Anand-Kumar} and is omitted here for brevity. Note that the result is the same as the DMT performance of the NAF protocol when infinite length waveforms are used.

\subsection{Asynchronous NAF with Finite Length Waveforms}
By pursuing the same procedure as we presented for the NAF protocol in Section \ref{Section_NAF_finite_u}, the received signal model in both phases is given by
\begin{equation}
  \underline{y} = \mathbf{H}\underline{x}+\underline{z},
\end{equation}
where
\begin{align*}
  \underline{x} =& \underline{x}'_0,\\
  \underline{y} =& \left[\big(\underline{y}_{d,0}'\big)^T,\underline{y}_{d,1}^T,
  \underline{y}_{d,2}^T,\ldots, \underline{y}_{d,M}^T\right]^T,\\
  \underline{z} =& \left[\big(\underline{z}_{d,0}'\big)^T, \underline{c}_1^T + \underline{z}_{d,1}^T, \underline{c}_2^T + \underline{z}_{d,2}^T,\ldots,\underline{c}_M^T + \underline{z}_{d,M}^T\right]^T,\\
  \mathbf{H} =& \left[h_0(\mathbf{\Gamma}'_{0,0})^T, \mathbf{G}^T\right]^T.
\end{align*}
In the above equation $\mathbf{G} = [\mathbf{G}_1^T, \ldots,\mathbf{G}_M^T] ^T$, where
\begin{align*}
  \mathbf{G}_i &= \sum_{j=1}^Mh_jg_j\mathbf{\Gamma}_{i,j}\mathbf{A}_j\mathbf{\Gamma}'_{0,0}\\
  \underline{c}_i & = \sum_{j=1}^Mh_j\mathbf{\Gamma}_{i,j}\mathbf{A}_j\underline{z}_{r_j}.
\end{align*}

The covariance matrix of the noise vector $\underline{z}$ is calculated as
\begin{equation}
  \mathbf{\Phi} = \sigma^2_d \left[
  \begin{array}{cc}
    \mathbf{\Gamma}'_{0,0} & \mathbf{0}_{p \times Mq} \\
    \mathbf{0}_{Mq \times p} & \mathbf{C}\\
   \end{array}\right],
\end{equation}
where $\mathbf{C} = [\mathbf{C}_{i,j}], \, i,j=1,2,\ldots,M$, and
\begin{equation*}
 \mathbf{C}_{i,j} = \mathbf{\Gamma}_{i,j}+\frac{\sigma_r^2}{\sigma_d^2}\sum_{k=1}^M|h_k|^2
  \mathbf{\Gamma}_{i,k}\mathbf{A}_k\mathbf{\Gamma}'_{0,0}\mathbf{A}_k^\dagger
  \mathbf{\Gamma}_{j,k}^\dagger.
\end{equation*}

Define
\begin{align}
 \mathbf{\Xi} \triangleq& \left[
  \begin{array}{cccc}
    \mathbf{\Gamma}_{1,1} & \mathbf{\Gamma}_{1,2} & \ldots & \mathbf{\Gamma}_{1,M}\\
    \mathbf{\Gamma}_{2,1} & \mathbf{\Gamma}_{2,2} & \ldots & \mathbf{\Gamma}_{2,M}\\
    \vdots & \vdots & \ldots & \vdots\\
    \mathbf{\Gamma}_{M,1}^\dagger &  \mathbf{\Gamma}_{M,2}^\dagger & \ldots & \mathbf{\Gamma}_{M,M}
   \end{array}\right],\\
   \mathbf{\Sigma} \triangleq& \left[h_1g_1\mathbf{A}_1^T,h_2g_2\mathbf{A}_2^T,\ldots,h_Mg_M\mathbf{A}_M^T\right]^T.
\end{align}
One can check that
\begin{align}
  \mathbf{G} &= \mathbf{\Xi\Sigma}\mathbf{\Gamma}'_{0,0}\\
  \mathbf{C} &= \left(\mathbf{\Xi}\texttt{diag}\{\hat{\mathbf{A}}_1,\ldots,\hat{\mathbf{A}}_M\}
  +\mathbf{I}_{Mq}\right) \mathbf{\Xi},
\end{align}
where $\hat{\mathbf{A}}_i = \frac{\sigma_r^2}{\sigma_d^2}|h_i|^2\mathbf{A}_i\mathbf{\Gamma}'_{0,0}
\mathbf{A}_i^\dagger$. Hence, $\mathbf{C}^{-1}$ exists if and only if $\mathbf{\Xi}^{-1}$ exists. According to Proposition \ref{Prop_det_Xi_k}, if the shaping waveforms $\psi_i(t), i=0,\ldots, M$ are designed properly, $\mathbf{\Xi}$ is positive definite and $\mathbf{\Xi}^{-1}$ exists. $\mathbf{\Gamma}'_{0,0}$ is also a full rank matrix with bounded positive real eigenvalues (see \cite{Robert-Gray}). Therefore, $\mathbf{\Phi}^{-1}$ is
given by
\begin{equation}
 \mathbf{\Phi}^{-1} = \frac{1}{\sigma_d^2}~\texttt{diag}\{(\mathbf{\Gamma}'_{0,0})^{-1}, \mathbf{C}^{-1}\}.
\end{equation}

By pursuing the same procedure as that of the NAF protocol, for high values of SNR, the mutual information between the source and the destination is obtained as
\begin{align}\nonumber
 I(\underline{x};\underline{y}) & \doteq \log (1+\rho|h_0|^2)^{p-q} \left(1+\rho|h_0|^2 +\sum_{i=1}^M|h_ig_i|^2\right)^q\\
 & \doteq \left[(p-q)(1-\alpha_0)^+ + q\max\{1-\alpha_0, 1-\beta\}^+\right]\log \rho.
\end{align}
where $\alpha_0 \triangleq -\frac{\log|h_0|^2}{\log \rho}$, $\beta_i \triangleq -\frac{\log|h_ig_i|^2}{\log \rho}$, and $\beta = \min_{i\ge 1} \beta_i$. For the rate $R = r\log \rho$, the outage probability is given
by
\begin{equation}\nonumber
P_{\mathcal{O}}(r\log \rho) = Pr(I(\underline{x};\underline{y})< \ell r\log \rho) \doteq \rho^{-d^*(r)},
\end{equation}
where for $\alpha_i,\beta_i \ge 0,\, \forall i \in \{0,1,\ldots,M\}$
\begin{equation}
  d^*(r) = \inf_{(p-q)(1-\alpha_0)^+ + q\max\{1-\alpha_0,1-\beta \}^+< \ell r} \alpha_0+M\beta.
\end{equation}
Clearly, $\inf (\alpha_0+M\beta)$ occurs when $0 \le \alpha_0, \beta \le 1$. Hence,
\begin{equation}
  d^*(r) = \inf_{(p-q)\alpha_0 + q\min\{\alpha_0,\beta\} > p-\ell r} \alpha_0+M\beta.
\end{equation}
As can be seen the optimum diversity gain in this case is the same as that of the OAF protocol when infinite length waveforms are used given in \eqref{Eqn_d*(r)_OAF_Infinite_p}. We simply give the final result for the optimum value of $\kappa$ at each multiplexing gain in the following theorem.
\begin{theorem}
For a finite value of $u$, the DMT performance of the OAF protocol over the underlying asynchronous network when $\kappa$ varies to maximize the diversity gain is given by.
\begin{equation}\nonumber
  d(r) = M(1-2r)^+ + (1-r)^+.
\end{equation}
The best DMT for $0 \le r \le \frac{1}{2}$ is achieved when $\kappa = \frac{M+1}{M}$. For $\frac{1}{2}\le r \le 1$, the best DMT is achieved when the source transmits alone.
\end{theorem}

The proof is similar to the proof of Theorem \ref{Theorem_OAF_infinite_p} which is given in \cite{Elia-Vinodh-Anand-Kumar} and is omitted here for brevity. Note that  the result is similar to the DMT result of the asynchronous network with infinite length shaping waveforms presented in Section \ref{Subsec_OAF_infinite_u} and equivalently similar to the DMT result of the corresponding synchronous network.
\begin{figure}
\centering
\includegraphics[width=9 cm , height = 7cm]{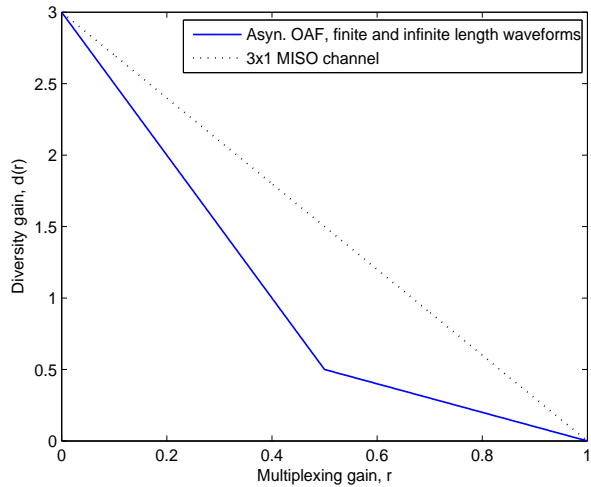}
\caption{The DMT performances of the asynchronous NSDF protocol for both finite and infinite length shaping waveforms and optimum values of $\kappa$.}\label{OAF_OptimumK_Figure}
\end{figure}
Fig. \ref{OAF_OptimumK_Figure} illustrates the DMT curves of the OAF protocol over the two relay asynchronous cooperative network for both cases of using finite length shaping waveforms and using infinite length shaping waveforms when $\kappa$ is chosen to maximize the diversity gain at each multiplexing gain $r$. As can be seen, the OAF protocol over the underlying network performs the same as the corresponding synchronous protocol for both scenarios.

\section{Discussion and Conclusion}\label{Discussion}
In this paper, we examined the DMT performances of the NSDF, OSDF, NAF, and OAF relaying protocols over a general two-hop \textit{asynchronous} cooperative relay network containing one source node, one destination node, and $M$ parallel relay nodes. To model the asynchronism, we assumed the nodes send PAM signals asynchronously wherein information symbols are linearly modulated by a shaping waveform. We analyze the DMT of the system from both theoretical and practical points of views. In the former, we consider the case that all transmitters use shaping waveforms with infinite time support resulting in a communication over a strictly limited bandwidth. We showed that asynchronism in this case preserves the DMT performances of the system for all the aforementioned protocols. In the latter where finite length shaping waveforms (as in practice) are used, the communication is carried out over a spectral mask which is not strictly limited in the frequency domain and its tails go to infinity from both sides. We showed that in this case the asynchronism helps to improve the DMT performance in the NSDF, OSDF, and NAF protocols, while preserves the DMT in the OAF protocol.

\subsection{Comparison of the DMT Performances of the Protocols}
\begin{figure*}[htp]
  \centering
  \includegraphics[width = 16cm, height = 11 cm]{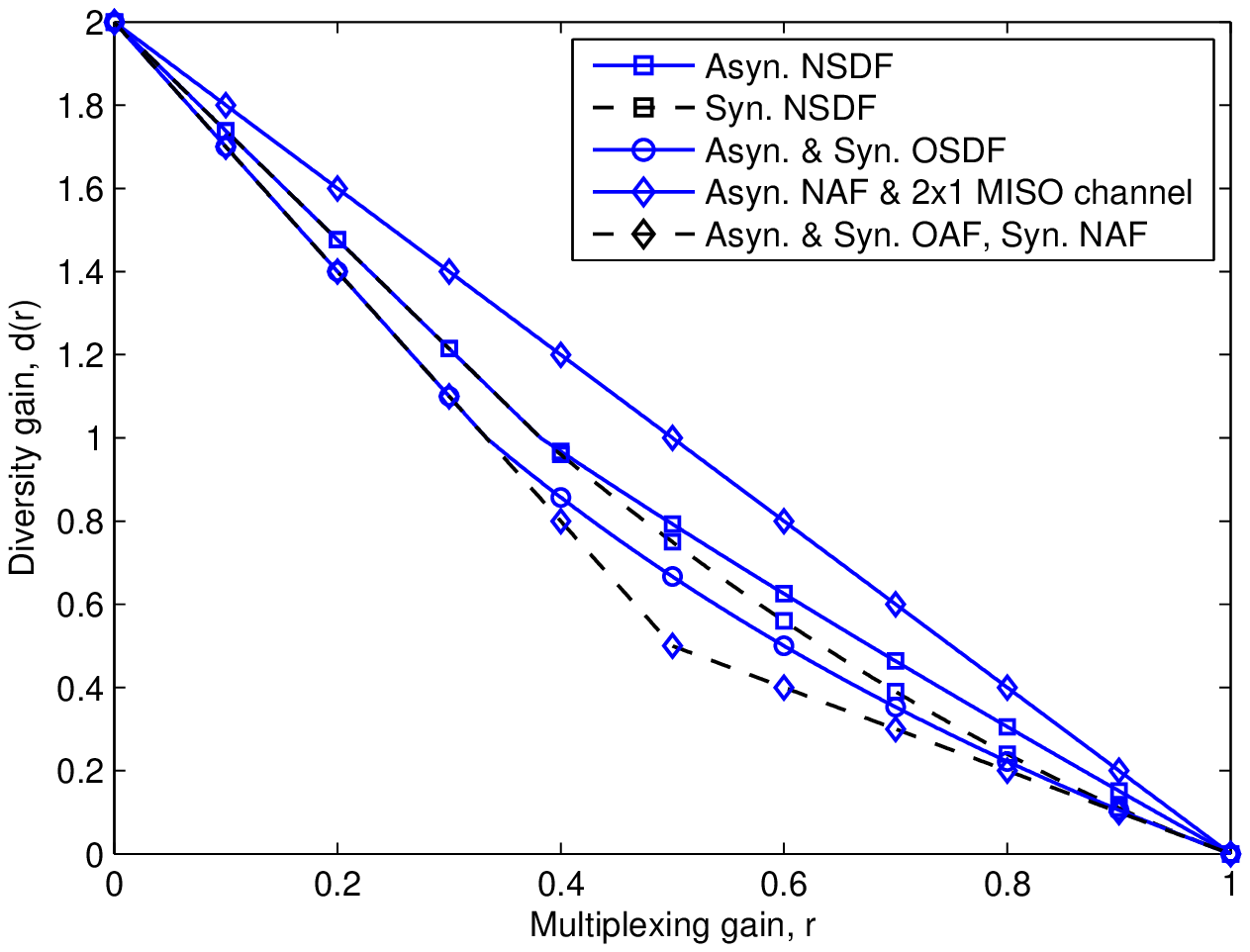}
  \caption{The DMT performances of the asynchronous protocols and the corresponding synchronous counterparts in a single relay network.}\label{DMT_Comparison_SingleRelay_Figure}
  \centering
  \includegraphics[width = 16cm, height = 11 cm]{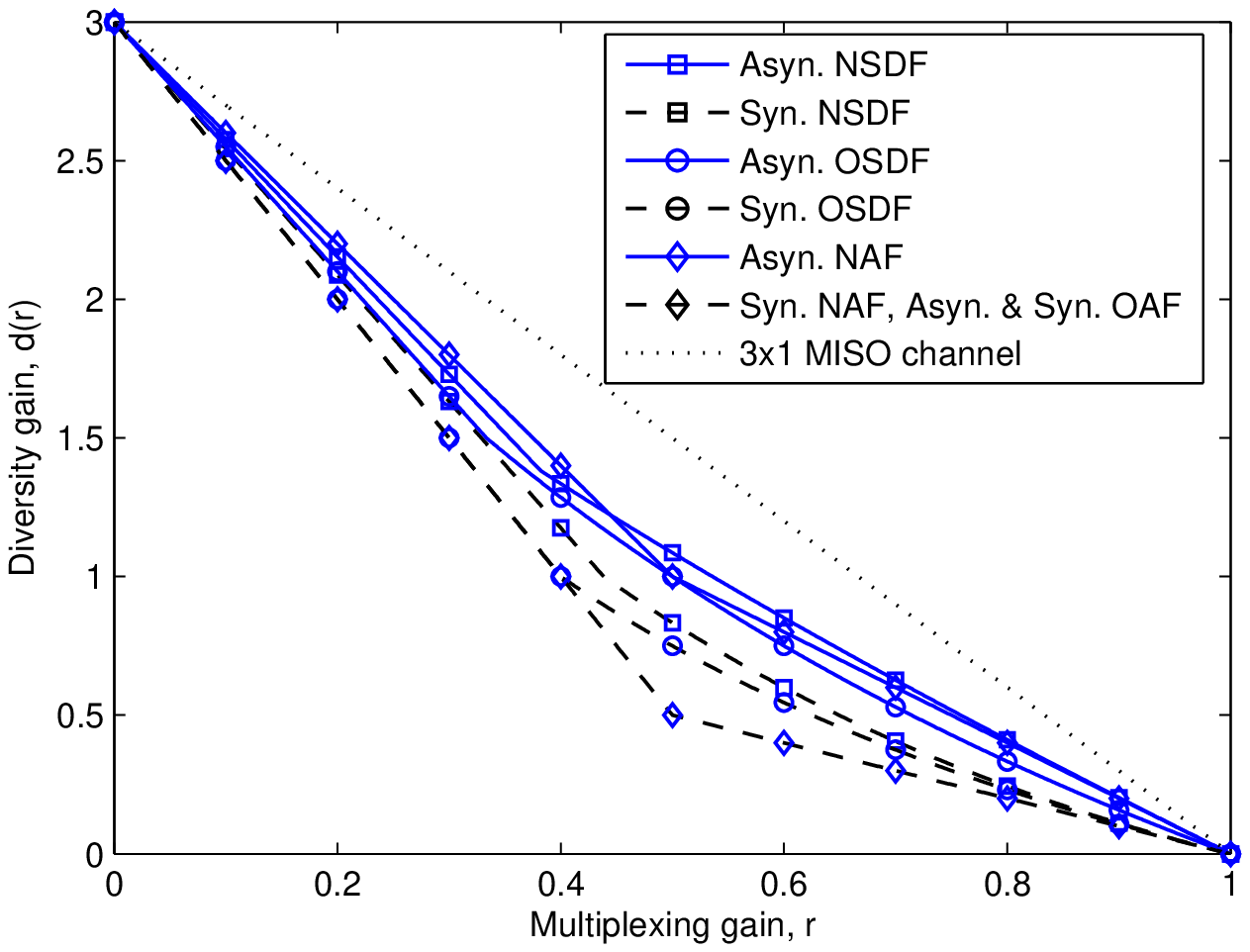}
\caption{DMT performance of the asynchronous protocols and the corresponding synchronous counterparts in a two relay
network.}\label{DMT_Comparison_TWoRelay_Figure}
\end{figure*}
Figures \ref{DMT_Comparison_SingleRelay_Figure} and \ref{DMT_Comparison_TWoRelay_Figure} demonstrate the DMT performances of the discussed relaying protocols over the single relay and the two relay cooperative networks for both cases of using finite length and using infinite length shaping waveforms. Note that for all protocols, the DMT performance of the underlying asynchronous network with infinite length shaping waveforms is the same as that of the corresponding synchronous network. As shown, except in OAF where both scenarios show the same DMT performances, in all other scenarios, asynchronous protocols with finite length shaping waveforms outperform the corresponding counterparts. In the single relay network, the asynchronous NAF with finite length waveforms achieves the same DMT performance as that of the $2 \times 1$ MISO channel. However, it only shows the best DMT performance in low multiplexing gain regime over the two relay network. In the high multiplexing gain regime, the asynchronous NSDF with finite length waveforms yields the best performance. One can check that by increasing the number of helping nodes ($M \ge 3$), this protocol becomes superior throughout the range of the multiplexing gain, while asynchronous NAF settles at the third place after the asynchronous OSDF protocol both with finite length waveforms. Note that the extra diversity gain when the shaping waveforms are of limited time support is at the expense of a bandwidth expansion at high values of SNR.

\subsection{Where Do the Gains Come From?}
The main objective of this work is to show that the asynchronism does not diminish the DMT performance of a general two-hop cooperative network under the aforementioned relaying protocols. Moreover, when a practical cooperative network is considered wherein PAM signals with finite length shaping waveforms are used, even better diversity gains can be achieved at the presence of the asynchronism. This gain is due to the fact that the communication in this case is carried out over a spectral mask with tails spanning over the entire frequency axis. This causes the mutual information between the source and the destination to be similar to that of a parallel channel with the number of parallel branches equal to the number of links that carry independent codewords. For example, in DF type protocols where all links carry independent Gaussian codewords, the number of parallel links is equal to the number of transmitting nodes. In contrast, in the OAF protocol where all nodes carry correlated signals, the resulted mutual information of the asynchronous channel is the same as that of the corresponding synchronous channel and no parallel links appear. Note that the asynchronism is a critical factor to extract this gain from such channels. One can easily check that if the system is fully synchronous and the same shaping waveforms with a finite time support are used, this gain is not revealed. This clears the advantage of asynchronous signaling over such channels.

\subsection{Shaping Waveforms}
The results of this work are applied to regular shaping waveforms used in theoretical analysis (e.g., the ``sinc'' and the ``raised-cosine'' waveforms). The truncated versions of such waveforms are extensively used in practice. One can easily see that the required condition in equation \eqref{gis_condition} is held when all nodes use shaping waveforms with infinite time support. On the other hand, if all the waveforms have a limited time support, this condition barely holds when the nodes are randomly asynchronous.

\subsection{Practical Implementation}
In practice, we propose using OFDM (inverse discrete Fourier Transform (DFT) at the transmitters and DFT at the receivers) to implement the asynchronous protocols. It can be shown that the same DMT performances can be achieved in the limit of the codeword's length. In this case, a DMT achieving space-times code designed for synchronous cooperative networks \cite{Elia-Vinodh-Anand-Kumar} can also achieve the DMT of the corresponding asynchronous network.

Although it was assumed that the asynchronous delays are less than a symbol interval, the results are still held in the limit of the codewords' length when the delays are arbitrary finite random variables. In this case, one can discard a few samples from both sides of a received frame or increase the length of the cyclic prefix symbols if OFDM is used to adjust the remaining asynchronism among the nodes to be less than a symbol interval. Since the number of the discarded symbols is finite, they do not affect the maximum multiplexing gain for large length codewords.

\appendices

\section{Proof of Lemma \ref{Lemma_Xi_Expression}}\label{Apx_Proof_Lemma_Xi_Expression}
Let $\underline{\psi}(t) \triangleq [\psi_0(t),\psi_1(t-\tau_{1,0}),\ldots,\psi_{m}(t-\tau_{m,0})]$ and $\underline{\psi}_{\omega}(t) = \sum_{v=-u}^u \underline{\psi}(t-vT_s)e^{-\xi \omega v}$. One can check that
\begin{align}\nonumber
  \mathbf{\Gamma}(\omega)&= \int_{-\infty}^{\infty}\underline{\psi}^\dagger(t)\underline{\psi}_{\omega}(t) dt\\ \nonumber
&= \sum_{v=-u}^u\int_{-\infty}^{\infty}\underline{\psi}^\dagger(t)\underline{\psi}(t-vT_s)e^{-\xi\omega v} dt\\
&= \sum_{v=-u}^{-1}\mathbf{A}_v+\mathbf{A}_0+\sum_{v=1}^{u}\mathbf{A}_v,
\end{align}
where $\mathbf{A}_v = \int_{-\infty}^{\infty}\underline{\psi}^\dagger(t)\underline{\psi}(t-vT_s)e^{-\xi\omega v}dt$. We have,
\begin{align}\nonumber
\sum_{v=-u}^{-1}\mathbf{A}_v &= \sum_{v=-u}^{-1} \int_0^{(u+v+1)T_s}\underline{\psi}^\dagger(t)\underline{\psi}(t-vT_s)e^{-\xi\omega v}dt\\ \nonumber
& = \sum_{v=1}^u\sum_{n=0}^{u-v} \int_0^{T_s} \underline{\psi}^\dagger(t+nT_s)\underline{\psi}(t+(n+v)T_s)e^{\xi\omega v}dt\\ \nonumber
& = \sum_{n=0}^{u}\int_0^{T_s} \underline{\psi}^\dagger(t+nT_s)\sum_{v=1}^{u-n}\underline{\psi}(t+(n+v)T_s)e^{\xi\omega v}dt.
\end{align}
\begin{align}\nonumber
\mathbf{A}_0 &= \int_{0}^{(u+1)T_S}\underline{\psi}^\dagger(t)\underline{\psi}(t) dt\\ \nonumber
&=\sum_{n=0}^u \int_{0}^{T_S} \underline{\psi}(t+nT_s)^\dagger\underline{\psi}(t+nT_s) dt.
\end{align}
\begin{align}\nonumber
\sum_{v=1}^{u}\mathbf{A}_v &= \sum_{v=1}^{u} \int_{vT_s}^{(u+1)T_s}\underline{\psi}^\dagger(t)\underline{\psi}(t-vT_s) e^{-\xi\omega v}dt\\ \nonumber
&= \sum_{v=1}^u\sum_{n=v}^{u} \int_0^{T_s} \underline{\psi}^\dagger(t+nT_s)\underline{\psi}(t+(n-v)T_s)e^{-\xi\omega v}dt\\ \nonumber
&=\sum_{n=0}^{u} \int_0^{T_s} \underline{\psi}^\dagger(t+nT_s)\sum_{v=1}^u\underline{\psi}(t+(n-v)T_s)e^{-\xi\omega v}dt.
\end{align}
$\mathbf{\Gamma}(\omega)$ can be rewritten as follows.
\begin{align*}
 \mathbf{\Gamma}(\omega) & = \sum_{n=0}^u\int_{0}^{T_s}\underline{\psi}^\dagger(t+nT_s)e^{-\xi\omega n}\Big[\underline{\psi}(t+nT_s)e^{\xi\omega n}+\sum_{v=1}^{u-n}\underline{\psi}(t+(v+n)T_s)e^{\xi\omega (v+n)}+\\
&\hspace{7.7cm}\sum_{v=1}^{n}\underline{\psi}(t+(n-v)T_s)e^{\xi\omega (n-v)}\Big]dt \\
& = \int_{0}^{T_s}\left[\sum_{n=0}^u\underline{\psi}(t+nT_s)e^{\xi\omega q}\right]^\dagger\sum_{v=0}^u\underline{\psi}(t+vT_s)e^{\xi\omega v} dt.
\end{align*}
This concludes the proof.\hfill $\blacksquare$

\section{Shift Property of the DTFT for Non-Integer Delays}\label{Apx_Shift_Property_of_DFT}
\begin{lemma}
  let $x(t)$ be a signal with a limited bandwidth $W$. $x(n)$ and $\hat{x}(n), \, n\in \mathbb{Z}$ are two sequences of samples of this signal at $t=nT_s$ and $t=nT_s+\tau$, respectively. $X(\omega)$ and $\hat{X}(\omega)$ are defined as the DTFT of these two sequences. If the sampling frequency is chosen according to the Nyquist sampling Theorem, i.e., $W \le \frac{1}{2T_s}$, the shift property of the DTFT is held for any real value of $\tau$ and we get
  \begin{equation*}
    \hat{X}(\omega) = e^{\xi\omega\hat{\tau}},
  \end{equation*}
  where $\hat{\tau} = \frac{\tau}{T_s}$.
\end{lemma}
\begin{proof}
  Since $x(t)$ is bandlimited, it can be reconstructed from its samples if $W \le \frac{1}{2T_s}$ as follows.
  \begin{equation*}
    x(t) = \sum_n x(n)\texttt{sinc}\left(\frac{t-nT_s}{T_s}\right),
  \end{equation*}
  where $\texttt{sinc}(x)= \frac{\sin \pi x}{\pi x}$. W have
  \begin{align*}
    \hat{X}(\omega) &= \sum_k \hat{x}(k)e^{-\xi\omega k}\\
    &= \sum_k\sum_n x(n)\texttt{sinc}\left(\frac{(k-n)T_s+\tau}{T_s}\right)e^{-\xi\omega k}\\
    &= \sum_n x(n)\sum_k \texttt{sinc}\left(k-n+\hat{\tau}\right)e^{-\xi\omega k}\\
    &= e^{\xi\omega\hat{\tau}} \sum_n x(n)e^{-\xi\omega n}\\
    &= e^{\xi\omega\hat{\tau}}X(\omega),
  \end{align*}
where the second last equality is due to the fact that the DTFT of $\texttt{sinc}(n+a)$ is equal to $e^{\xi\omega a}$ for all real $a$. This concludes the proof.
\end{proof}

\section{Proof of Proposition \ref{Prop_RankofGammaj}}\label{Appendix_Proof_Prop_RankofGammaj}
Define $\tilde{\mathbf{\Gamma}}_j$ of size $N \times N, \, N>2q$ as in \eqref{Eqn_tildeGamma_j}.
\begin{equation}\label{Eqn_tildeGamma_j}
  \tilde{\mathbf{\Gamma}}_j = \left[
  \begin{array}{ccccccccccccc}
    \gamma_j(0) & \cdots & \gamma_j(-q+1) & 0 &0 & \ldots & 0 & \gamma_j(q-1) & \ldots & \gamma_j(1)\\
    \ddots & \ddots & \ddots & \ddots & \ddots & \ddots & \ddots & \ddots & \ddots & \ddots\\
    \gamma_j(q-1) & \cdots & \gamma_j(0) & \cdots & \gamma_j(-q+1)& 0 & \ldots & 0 & \cdots & 0\\
    \ddots & \ddots & \ddots & \ddots & \ddots & \ddots & \ddots & \ddots & \ddots\\
    0 & \cdots & 0 & \cdots & 0 & \gamma_j(q-1) & \cdots & \gamma_j(0) & \cdots & \gamma_j(-q+1)\\
    \ddots & \ddots & \ddots & \ddots & \ddots & \ddots & \ddots & \ddots & \ddots\\
    \gamma_j(-1) & \cdots & \gamma_j(-q+1) & 0 &   \cdots & \cdots & 0 & \gamma_j(q-1) & \cdots & \gamma_j(0)\\
  \end{array}\right].
\end{equation}
$\tilde{\mathbf{\Gamma}}_j$ is the circular convolution matrix of the sequence $\hat{\underline{\gamma}}_j = [\gamma_j(0),\ldots,\gamma_j(q-1),0,\ldots,0, \gamma_j(-q+1),\ldots,\gamma_j(-1)]$ of length $N$. Hence, it can be decomposed as $\tilde{\mathbf{\Gamma}}_j = \mathbf{U}_N\mathbf{\Lambda}_j\mathbf{U}_N^\dagger$,
where $\mathbf{U}_N$ is the discrete Fourier transform (DFT) matrix of dimension $N$ defined as
\begin{equation}
  \mathbf{U}_N(i,j) = \frac{1}{\sqrt{N}}e^{-\xi\frac{2\pi(i-1)(j-1)}{N}}, i,j=1,2,\ldots,N,
\end{equation}
and $\mathbf{\Lambda}_j$ is a diagonal matrix containing the DFT elements of the vector $\hat{\underline{\gamma}}_j$ on its main diagonal. The $k$-th diagonal entry of this matrix is given by
\begin{equation}
  \mathbf{\Lambda}_j(k,k) = \sum_{n=0}^{N-1}\hat{\gamma}_j(k)e^{-\xi\frac{2\pi}{N}kn},
\end{equation}
where $\hat{\gamma}_j(k)$ is the $k$-th entry of $\underline{\hat{\gamma}}_j$. If $\psi_j(t)$ has a non-zero spectrum over the bandwidth $W$ and the sampling frequency $f_s=2W$ is chosen, the DFT vector of $\underline{\hat{\gamma}}_j$ does not have any deterministic zero. Hence, $\mathbf{\Lambda}_j$ and accordingly $\tilde{\mathbf{\Gamma}}_j$ are full rank matrices. Sine $\mathbf{\Gamma}_j$ is the top left sub matrix of $\tilde{\mathbf{\Gamma}}_j$, it is also a full rank matrix.

Let $M_f$ be the essential supremum $M_f = \texttt{ess sup} f$ of a real value function $f(x)$ which is defined as the smallest number $a$ for which $f(x)\le a$ except on a set of measure zero. Let $m_f$ be the essential infimum $m_f = \texttt{ess inf} f$ of a real value function $f(x)$ which is defined as the largest number $a$ for which $f(x)\ge a$ except on a set of measure zero. Let $\lambda_k,\, k=1,2,\ldots,q$ be the $k$-th eigenvalue of $\mathbf{\Gamma}_j$. It is proved in \cite{Robert-Gray} that if $\mathbf{\Gamma}_j$ is Hermitian
\begin{equation}\nonumber
  m_f \le \lambda_k \le M_f,
\end{equation}
whether or not
\begin{equation}\nonumber
  \max_k \lambda_k \le 2M_{|f|},
\end{equation}
where $f$ here is the DTFT function of the samples of the shaping waveform $\psi_j(t)$. Since $m_f, M_f$, and $M_{|f|}$ are bounded values for well-designed shaping waveforms, therefore, $\mathbf{\Gamma}_j$ is a full rank matrix with non-zero bounded eigenvalues for all $j\in\{0,1,\ldots,m\}$.
This concludes the proof. \hfill $\blacksquare$

\section{Proof of Lemma \ref{Lemma_dEm_NSDF_FixK}}\label{Proof_Lemma_dEm_NSDF_FixK}
Clearly, $\inf \sum_{i=1}^m\alpha_i$ occurs in the
region $0 \le \alpha_i \le 1, ~ i \in \{0,1,\ldots, m\}$.
Hence, we focus on this region to proceed the proof.
\begin{eqnarray}\nonumber
  &&(1-\alpha_0)^++\frac{q}{\ell}\sum_{i=1}^m (1-\alpha_i)^+< r \\ \nonumber
  &\Rightarrow& \alpha_0 + \frac{q}{\ell}\sum_{i=1}^m\alpha_i >
  1 + \frac{mq}{\ell}-r\\ \nonumber
  &\Rightarrow& \sum_{i=1}^m\alpha_i > \frac{\ell}{q}\left[(1-\alpha_0)
  +\frac{mq}{\ell}-r\right]\\ \nonumber
  &\Rightarrow& \sum_{i=1}^m \alpha_i > m + \frac{\ell}{q}(1-\alpha_0)-\frac{\ell}{q}r.
\end{eqnarray}
Hence
\begin{eqnarray}\nonumber
d_{E_m}(r) =
  \inf_{0 \le \alpha_0 \le 1} \alpha_0 + \max\left\{0,
  m + \frac{\ell}{q}(1-\alpha_0)-\frac{\ell}{q}r\right\}.~~~~
\end{eqnarray}
If $0 \le r \le \frac{mq}{\ell}$, then $(m +
\frac{\ell}{q}(1-\alpha_0)-\frac{\ell}{q}r) \ge 0$ for all $0 \le \alpha \le 1$. Hence,
\begin{eqnarray}\nonumber
  d_{E_m}(r) &=& \inf_{0 \le \alpha_0 \le 1} \left(m+1 + \frac{p}{q}(1-\alpha_0)-\frac{\ell}{q}r \right)\\ \nonumber
  &=& 1+m-\frac{\ell}{q}r,~~~ 0 \le r \le \frac{mq}{\ell}.
\end{eqnarray}
For $r \ge \frac{mq}{\ell}$, if $\left(m+
\frac{\ell}{q}(1-\alpha_0)-\frac{\ell}{q}r\right)\ge 0$, then
$\alpha_0 \le 1+\frac{mq}{\ell}-r$. In this
case, we have
\begin{eqnarray}\nonumber
  d_{E_m}(r) &=& \inf_{0 \le \alpha_0 \le 1+\frac{mq}{\ell}-r}
  m+1 + \frac{p}{q}(1-\alpha_0) - \frac{\ell}{q}r\\ \nonumber
  &=& 1+\frac{mq}{\ell}-r, ~~~ \frac{mq}{\ell} \le r \le 1.
\end{eqnarray}
In contrast, when $\alpha_0 >
1+\frac{mq}{\ell}-r$, we have
\begin{eqnarray}\nonumber
    d_{E_m}(r) &=& \inf_{1+\frac{mq}{\ell}- r \le \alpha_0  \le 1} \alpha_0\hspace{3 cm}\\ \nonumber
    &=& 1+\frac{mq}{\ell}-r, ~~~ \frac{mq}{\ell} \le r \le 1.
\end{eqnarray}
Hence, for $m \le \kappa+1$
\begin{equation}\nonumber
  d_{E_m}(r) = \left\{
  \begin{array}{ll}
    1+m-\frac{\ell}{q}r, & 0 \le r \le \frac{mq}{\ell}\\
    1+\frac{mq}{\ell}-r, & \frac{mq}{\ell} < r \le 1.
  \end{array}\right.
\end{equation}
For $m \ge \kappa+1$, $\frac{mq}{\ell} \ge 1$.
Thus
\begin{equation}\nonumber
  d_{E_m}(r) =  1+m-\frac{\ell}{q}r, ~~~ 0 \le r \le 1.
\end{equation}
This concludes the proof. \hfill $\blacksquare$

\section{Proof of Proposition \ref{Proposition_NSDF_1Relay}}\label{Proof_Proposition_NSDF_1Relay}
The outage probability is calculated as
\begin{align}\nonumber
  P_{\mathcal{O}}(R) &= Pr(E_0)P_{\mathcal{O} \mid E_0} + Pr(E_1)P_{\mathcal{O} \mid E_1}\\ \nonumber
  &\doteq \left\{
  \begin{array}{cc}
    \rho^{-[(1-\frac{\ell}{p}r)+(1-r)]}+\rho^{-(2-\frac{\ell}{q}r)}, & 0 \le r \le \frac{q}{\ell}\\
    \rho^{-[(1-\frac{\ell}{p}r)+(1-r)]}+\rho^{-(1+\frac{q}{\ell}
    -r)}, & \frac{q}{\ell} \le r \le \frac{p}{\ell}\\
    \rho^{-(1-r)}, & \frac{p}{\ell} \le r \le 1.\\
  \end{array}\right.
\end{align}
In each region, the term with the largest exponent of $\rho$ is dominant. We consider three distinct regions $0 \le r \le \frac{q}{\ell}$, $\frac{q}{\ell} \le r \le \frac{p}{\ell}$, and
$\frac{p}{\ell} \le r \le 1$ and evaluate the
diversity gain in each region. For $0 \le r \le \frac{q}{\ell}$,
\begin{align}\nonumber
  \text{If} ~~ \Big[\Big(1-\frac{\ell}{p}r\Big)+\left(1-r\right)\Big]\le \Big(2-\frac{\ell}{q}r\Big)
  &~~ \Rightarrow~ \frac{\ell}{p}+1 \ge \frac{\ell}{q} \\ \nonumber
  &~~\Rightarrow~ \kappa^2-\kappa-1 \le 0.
\end{align}
Hence assuming $\kappa \ge 1$, for $0 \le r \le \frac{q}{\ell}$ we have
\begin{equation}
  d^*(r) = \left\{
  \begin{array}{cc}
    \Big(1-\frac{\ell}{p}r\Big) + (1-r), & 1 \le \kappa \le \hat{\kappa}\\
    2\Big(1-\frac{\ell}{2q}r\Big), & \kappa \ge \hat{\kappa}
  \end{array} \right.
\end{equation}
where $\hat{\kappa} = \frac{1+\sqrt{5}}{2}$.

For $\frac{q}{\ell} \le r \le \frac{p}{\ell}$,
\begin{align} \nonumber
  \text{If} ~~ \Big[\Big(1-\frac{\ell}{p}r\Big)+(1-r)\Big]
  \le \Big(1+\frac{q}{\ell}-r\Big)&~\Rightarrow~ 1-\frac{\ell}{p}r \le \frac{q}{\ell}\\ \nonumber
  &~\Rightarrow~ r \ge \frac{p^2}{\ell^2}.
\end{align}
Clearly $\frac{p^2}{\ell^2} \le \frac{p}{\ell}$. Moreover,
\begin{equation}\nonumber
  \text{if}~~ \frac{p^2}{\ell^2} \ge \frac{q}{\ell} ~\Rightarrow~ p^2 \ge q\ell ~ \Rightarrow~ \kappa^2 - \kappa - 1 \ge 0.
\end{equation}
Hence, if $1 \le \kappa \le \hat{\kappa}$, then $\frac{p^2}{\ell^2} \le \frac{q}{\ell}$ and we have
\begin{eqnarray}\nonumber
  d^*(r) = \Big(1-\frac{\ell}{p}r\Big) + (1-r), ~~~ \frac{q}{\ell} \le r \le \frac{p}{\ell}.
\end{eqnarray}
However, if $\kappa \ge \hat{\kappa}$, then $\frac{p^2}{\ell^2} \ge \frac{q}{\ell}$ and  therefore
\begin{equation}\nonumber
  d^*(r) = \left\{
  \begin{array}{cc}
    1+\frac{q}{\ell}-r, & \frac{q}{\ell} < r \le \frac{p^2}{\ell^2}\\
    \Big(1-\frac{\ell}{p}r\Big) + (1-r), & \frac{p^2}{\ell^2} < r \le \frac{p}{\ell}.
  \end{array}\right.
\end{equation}
For $\frac{p}{\ell} \le r \le 1$, it is clear that
\begin{equation}\nonumber
 d^*(r) = 1-r.
\end{equation}

By combining the results of all the regions, we have
\begin{equation}\nonumber
d^*(r) = \Big(1-\frac{\ell}{p}r\Big)^+ + \left(1-r\right),~~~ 0 \le r\le 1,
\end{equation}
when $1 \le \kappa \le \hat{\kappa}$ and
\begin{equation}\nonumber
d^*(r) = \left\{
      \begin{array}{cc}
       2(1-\frac{\ell}{2q}r), & 0 \le r \le \frac{q}{\ell}\\
       1+\frac{q}{\ell}-r, & \frac{q}{\ell} < r \le \frac{p^2}{\ell^2}\\
       (1-\frac{\ell}{p}r) + (1-r), & \frac{p^2}{\ell^2} < r \le \frac{p}{\ell}\\
       1-r, & \frac{p}{\ell} \le r \le 1,
    \end{array}\right.
\end{equation}
when $\kappa \ge \hat{\kappa}$.
This concludes the proof of the first part of the Theorem.

For $r \le \frac{q}{\ell}$, the maximum diversity gain is achieved when $\kappa = \hat{\kappa} = \frac{1+\sqrt{5}}{2}$. If the optimum value of $\kappa$ is chosen for this region, we have
\begin{equation}\nonumber
 r \le \frac{q}{\ell} = \frac{1}{1+\hat{\kappa}}.
\end{equation}
The corresponding diversity gain in this region is given by
\begin{align}\nonumber
  d^*(r) &= \Big(1-\frac{\ell}{p}r\Big) + (1-r)\\
  &= 2-\frac{2\hat{\kappa}+1}{\hat{\kappa}}r, ~~~ 0 \le r \le \frac{1}{1+\hat{\kappa}}.
\end{align}

For a specific $r > \frac{1}{1+\hat{\kappa}}$, the maximum diversity gain  is achieved when $r =
\frac{p^2}{\ell^2}$. In this case,
\begin{equation}\nonumber
 r = \frac{p^2}{\ell^2} = \frac{\kappa^2}{(1+\kappa)^2}.
\end{equation}
Hence, for $r > \frac{1}{1+\hat{\kappa}}$ and $\kappa > 1$
\begin{equation}
 \kappa(r) = \frac{\sqrt{r}}{1-\sqrt{r}}.
\end{equation}
The corresponding diversity gain is given by
\begin{align}\nonumber
 d(r) &= 1+\frac{q}{\ell}-r\\ \nonumber
 &= 1+\frac{1}{\kappa(r)+1}-r\\
 &= 2-\sqrt{r}-r.
\end{align}

By combining the results of all the regions we have
\begin{equation}\nonumber
 d^*(r) = \left\{
\begin{array}{cc}
 1- (1+\frac{1}{\hat{\kappa}})r + (1-r), & 0 \le r \le \frac{1}{1+\hat{\kappa}}\\
 (1-\sqrt{r})+(1-r), & \frac{1}{1+\hat{\kappa}} \le r \le 1.
\end{array}\right.
\end{equation}
 The optimum $\kappa$ corresponding to each $r$ is given by
\begin{equation}\nonumber
 \kappa(r) = \left\{
 \begin{array}{cc}
 \hat{\kappa}, & 0 \le r \le \frac{1}{1+\hat{\kappa}}\\
 \frac{\sqrt{r}}{1-\sqrt{r}}, & \frac{1}{1+\hat{\kappa}} \le r \le 1.
 \end{array}\right.
\end{equation}
This concludes the proof. \hfill$\blacksquare$

\section{Proof of Theorem \ref{Theorem_NSDF_MRelay}}\label{Proof_Theorem_NSDF_MRelay}
In asynchronous NSDF protocol, if $M \le \kappa+1$
\begin{equation}\nonumber
d_{E_M}(r)= \left\{
\begin{array}{cc}
 1+M-\frac{\ell}{q}r, & 0 \le r \le \frac{Mq}{\ell}\\
 1+\frac{Mq}{\ell}- r, & \frac{Mq}{\ell} \le r \le 1,
\end{array}\right.
\end{equation}
else if $M \ge \kappa+1$
\begin{equation}\nonumber
d_{E_M}(r)=
1+M-\frac{\ell}{q}r,~~~ 0 \le r \le  1.
\end{equation}
In addition,
\begin{equation}\nonumber
 Pr(E_M) = \left\{
\begin{array}{cc}
 1, & 0 \le r \le \frac{p}{\ell}\\
 0, & \frac{p}{\ell} < r \le 1,
\end{array}\right.
\end{equation}

Let $b_m(r),~ m=0, \ldots, M$ be the negative exponent of $\rho$ in the expression $Pr(E_m)\rho^{-d_{E_m}}$ when $\rho \rightarrow \infty$, i.e., $Pr(E_m)\rho^{-d_{E_m}} \doteq \rho^{-b_m(r)}$. The outage probability at high values of SNR is given by
\begin{equation}\nonumber
  P_\mathcal{O} \doteq \sum_{i=0}^M \rho^{-b_m(r)}.
\end{equation}
If $d_{M-1}^*(r)$ is the DMT performance of the NSDF protocol over a cooperative network containing $M-1$ relays, $d_M^*(r)$ can be expressed as follows.
\begin{equation}\nonumber
 d_M^*(r) = \min\Big\{\Big(1-\frac{\ell}{p}r\Big)+d_{M-1}^*(r),b_M(r)\Big\},
\end{equation}
which is simplified as follows.

If $\kappa \le M-1$, then $d_M^*(r)$ is given by
\begin{equation}\nonumber
d_M^*(r) = \min\Big\{\Big(1-\frac{\ell}{p}r\Big)+d_{M-1}^*(r),1+M-\frac{\ell}{q}r\Big\},~~ 0 \le r \le \frac{p}{\ell}.
\end{equation}
Else for $\kappa \ge M-1$, $d_M^*(r)$ is given by
\begin{equation}\nonumber
d_M^*(r) =
\left\{
\begin{array}{c}
\min\{(1-\frac{\ell}{p}r)+d_{M-1}^*(r),1+M-\frac{\ell}{q}r\},~~ 0 \le r \le \frac{Mq}{\ell}\\
\min\{(1-\frac{\ell}{p}r)+d_{M-1}^*(r),1+\frac{Mq}{\ell}-r\},~~ \frac{Mq}{\ell} \le r \le \frac{p}{\ell}.
\end{array}\right.
\end{equation}
For $\frac{p}{\ell} \le r \le 1$, the source node transmits alone and
\begin{equation}\nonumber
 d_M^*(r) = 1-r.
\end{equation}

It can be seen that
\begin{align*}
 \text{If}~~~~~  1+M-\frac{\ell}{q}r \le b_0(r)
 &~\Rightarrow~ 1+M-\frac{\ell}{q}r \le M\Big(1-\frac{\ell}{p}r\Big)+(1-r)\\ \nonumber
 &~\Rightarrow~ \kappa^2 - M\kappa -M \ge 0,
\end{align*}
Thus, for $\kappa \le \frac{M+\sqrt{M^2+4M}}{2}$, $b_M(r) \ge b_0(r)$ and the event $E_M$ does not determine the DMT performance of the system. Hence for $\kappa \le M-1$,
\begin{equation}
 d_M^*(r) = \Big(1-\frac{\ell}{p}r\Big)+d^*_{M-1}(r), ~~~ 0 \le r \le \frac{p}{\ell}.
\end{equation}
For $\kappa \ge M-1$, if $0 \le r \le \frac{Mq}{\ell}$,
\begin{equation}
 d_M^*(r) = \Big(1-\frac{\ell}{p}r\Big)+d^*_{M-1}(r), ~~~ 0 \le r \le \frac{Mq}{\ell}.
\end{equation}
In this region of $\kappa$, For $\frac{Mq}{\ell} \le r \le \frac{p}{\ell}$ we have
\begin{align}\nonumber
&b_0(r) = M\Big(1-\frac{\ell}{p}r\Big)+1-r,\\ \nonumber
&b_1(r) = (M-1)\Big(1-\frac{\ell}{p}r\Big)+1+\frac{q}{\ell}-r,\\ \nonumber
&b_2(r) = (M-2)\Big(1-\frac{\ell}{p}r\Big)+1+\frac{2q}{\ell}-r,\\ \nonumber
& \hspace{3.5 cm} \vdots\\ \nonumber
&b_{M-1}(r) = \Big(1-\frac{\ell}{p}r\Big)+1+\frac{(M-1)q}{\ell}-r,\\ \nonumber
&b_M(r) = 1+\frac{Mq}{\ell}-r.
\end{align}
It can be seen that if $r \le \frac{p^2}{\ell^2}$, then $b_M(r) \le b_{M-1}(r) \le b_{M-2}(r) \le \ldots \le b_0(r)$. Otherwise, $b_M(r) \ge b_{M-1}(r) \ge b_{M-2}(r) \ge \ldots \ge b_0(r)$. One can check that if $\kappa \ge \frac{M+\sqrt{M^2+4M}}{2}$, then $\frac{p^2}{\ell^2} \ge \frac{Mq}{\ell}$. Hence, for $\kappa \ge \frac{M+\sqrt{M^2+4M}}{2}$ we have
\begin{equation}
d_M^*(r) = \left\{
\begin{array}{cc}
 1+\frac{Mq}{\ell}-r, & \frac{Mq}{\ell} \le r \le \frac{p^2}{\ell^2}\\
 M(1-\frac{\ell}{p}r)+1-r, & \frac{p^2}{\ell^2} \le r \le \frac{p}{\ell}.
\end{array}\right.
\end{equation}
For $\kappa \le \frac{M+\sqrt{M^2+4M}}{2}$, $\frac{p^2}{\ell^2} \le \frac{Mq}{\ell}$ and the event $E_M$ does not affect the DMT performance. By combining the results we have
\begin{equation}\nonumber
  d_M^*(r) = \Big(1-\frac{\ell}{p}r\Big)+d^*_{M-1}(r), ~~~ 0 \le r \le \frac{p}{\ell}.
\end{equation}
when $\kappa \le \frac{M+\sqrt{M^2+4M}}{2}$, and
\begin{equation}\nonumber
d_M^*(r) = \left\{
\begin{array}{cc}
 \Big(1-\frac{\ell}{p}r\Big)+d^*_{M-1}(r), & 0 \le r \le \frac{Mq}{\ell}\\
 1+\frac{Mq}{\ell}-r, & \frac{Mq}{\ell} \le r \le \frac{p^2}{\ell^2}\\
 M(1-\frac{\ell}{p}r)+1-r, & \frac{p^2}{\ell^2} \le r \le \frac{p}{\ell},\\
 (1-r), & \frac{p}{\ell} \le r \le \frac{p+n}{\ell}.
\end{array}\right.
\end{equation}

when $\kappa \ge \frac{M+\sqrt{M^2+4M}}{2}$. This concludes the proof of the first part of the Theorem. The proof of the second part is similar to the proof of the second part of Proposition \ref{Proposition_NSDF_1Relay}. \hfill$\blacksquare$

\section{Proof of Theorem \ref{Proposition_OSDF_2Relay}}\label{Proof_Proposition_OSDF_2Relay}
 For $M=2$, if $1 \le \kappa < 2$, we have
 \begin{align}\nonumber
   d_{E_0(r)} &= \left\{
   \begin{array}{ll}
     1-\frac{\ell}{p}r, & 0 \le r \le \frac{p}{\ell}\\
     0, & \frac{p}{\ell} < r,
   \end{array}\right.\\ \nonumber
   d_{E_1(r)} &= \left\{
   \begin{array}{ll}
     2-\frac{\ell}{q}r, & 0 \le r \le \frac{q}{\ell}\\
     1+\frac{q}{p}-\frac{\ell}{p}r, & \frac{q}{\ell} \le r \le \frac{p}{\ell} \\
     0, & \frac{p}{\ell} < r,
   \end{array}\right.\\ \nonumber
   d_{E_2(r)} &= \left\{
   \begin{array}{ll}
     3-\frac{\ell}{q}r, & 0 \le r \le \frac{p}{\ell}\\
     0, & \frac{p}{\ell} < r,
   \end{array}\right.
 \end{align}
The outage probability in this region, $1 \le \kappa \le 2$, is given by
\begin{align}\nonumber
  P_{\mathcal{O}} &= \sum_{i=0}^2P_{\mathcal{O}\mid E_i}Pr(E_i) \doteq \left\{
  \begin{array}{cc}
    \rho^{-d_1(r)}, & 0 \le r \le \frac{q}{\ell}\\
    \rho^{-d_2(r)}, & \frac{q}{\ell} \le r \le \frac{p}{\ell}\\
    1, & \frac{p}{\ell} < r.
  \end{array}\right.
\end{align}
where $d_1(r) = \min\{3(1-\frac{\ell}{p}r), 3-\frac{\ell^2}{pq}r, 3-\frac{\ell}{q}r\}$, and $d_2(r) =
\min\{3(1-\frac{\ell}{p}r), 2(1-\frac{\ell}{p}r)+\frac{q}{p},
3-\frac{\ell}{q}r\}$.

Assume $0 \le r \le \frac{q}{\ell}$. Clearly $3-\frac{\ell^2}{pq}r < 3-\frac{\ell}{q}r$. Moreover, if
$3(1-\frac{\ell}{p}r) < 3- \frac{\ell^2}{pq}r$, then $\kappa < 2$. Hence,
\begin{equation}\nonumber
 d^*(r) = 3\Big(1-\frac{\ell}{p}r\Big), ~~~ 0 \le r \le \frac{q}{\ell}, ~~1 \le \kappa < 2
\end{equation}

Now consider $\frac{q}{\ell} < r \le \frac{p}{\ell}$. It can be seen that $ 3\left(1-\frac{\ell}{p}r\right)~\lesseqqgtr~ 3-\frac{\ell}{q}r$ if and only if $k \lesseqqgtr 3$.
Furthermore, if $3\left(1-\frac{\ell}{p}\right) < 2\left(1-\frac{\ell}{p}\right)+\frac{q}{p}$, then $r < \frac{p-q}{\ell}$. One can check that, if $\kappa < 2$, then $\frac{p-q}{\ell} < \frac{q}{\ell}$. Hence,
\begin{equation}\nonumber
 d^*(r) = 3\left(1-\frac{\ell}{p}r\right), ~~~ \frac{q}{\ell} \le r \le \frac{p}{\ell}, ~~ 1\le \kappa < 2.
\end{equation}
The cooperation is avoided whenever it is beneficial to do so.
\begin{equation}\nonumber
 \text{if}~~~ 1-r \ge 3\left(1-\frac{\ell}{p}r\right) ~\Rightarrow~ r \ge \frac{2p}{3\ell-p}.
\end{equation}
Thus for $1 \le \kappa < 2$,
\begin{equation}
 d^*(r) = \left\{
 \begin{array}{cc}
   3(1-\frac{\ell}{p}r), & 0 \le r \le \frac{2p}{3\ell-p}\\
   1-r, & \frac{2p}{3\ell-p} \le r \le 1.
 \end{array}\right.
\end{equation}

For $\kappa \ge 2$, $d_{E_0(r)}$ and $d_{E_1(r)}$ are the same as before. However, $d_{E_2(r)}$ is given by
\begin{equation}\nonumber
 d_{E_2(r)} = \left\{
\begin{array}{ll}
 3-\frac{\ell}{q}r, & 0 \le r \le \frac{2q}{\ell}\\
 1+\frac{2q}{p}-\frac{\ell}{p}r, & \frac{2q}{\ell} < r \le \frac{p}{\ell}\\
 0, & \frac{p}{\ell} < r.
\end{array}\right.
\end{equation}
The outage probability is given by
\begin{equation}\nonumber
  P_{\mathcal{O}} = \left\{
  \begin{array}{ll}
    \rho^{-d_1(r)}, & 0 \le r \le \frac{q}{\ell}\\
    \rho^{-d_2(r)}, & \frac{q}{\ell} \le r \le \frac{2q}{\ell}\\
    \rho^{-d_3(r)}, & \frac{2q}{\ell} \le r \le \frac{p}{\ell}\\
    1, & \frac{p}{\ell} < r.
  \end{array}\right.
\end{equation}
where $d_1(r) = \min\{3(1-\frac{\ell}{p}r),
3-\frac{\ell^2}{pq}r, 3-\frac{\ell}{q}r\}$, $d_2(r) = \min\{3(1-\frac{\ell}{p}r), 2(1-\frac{\ell}{p}r)+\frac{q}{p}, 3-\frac{\ell}{q}r\} $, and $d_3(r) = \min\{3(1-\frac{\ell}{p}r), 2(1-\frac{\ell}{p}r)+\frac{q}{p}, 1+\frac{2q}{p}-\frac{\ell}{p}r\}$. We focus on each of the
above regions for $r$ to calculate the diversity gain.

Assume $0 \le r \le \frac{q}{\ell}$. Clearly, $3-\frac{\ell^2}{pq}r < 3-\frac{\ell}{q}r$. Moreover, if
$\kappa \ge 2$, then $3-\frac{\ell^2}{pq}r \le 3(1-\frac{\ell}{p}r)$. Hence,
\begin{equation}
 d^*(r) = 3-\frac{\ell^2}{pq}r, ~~~ 0 \le r \le \frac{q}{\ell}, ~~ \kappa \ge 2.
\end{equation}

Now consider $\frac{q}{\ell} \le r \le \frac{2q}{\ell}$. One can see that $ 3(1-\frac{\ell}{p}) \lesseqqgtr
3-\frac{\ell}{q}r ~\Longleftrightarrow~ \kappa \lesseqqgtr 3$. On the other hand, if $ 3(1-\frac{\ell}{p}r) < 2(1-\frac{\ell}{p}r)+\frac{q}{p}$, then $1-\frac{\ell}{p}r< \frac{n}{p}$ which results in $r > \frac{p-q}{\ell}$.
It is clear that if $\kappa \ge 2 \Rightarrow \frac{p-q}{\ell} \ge \frac{q}{\ell}$. Therefor, for $2 \le \kappa \le 3$,
\begin{equation}
 d^*(r) = \left\{
 \begin{array}{ll}
  2(1-\frac{\ell}{p}r)+\frac{q}{p}, & \frac{q}{\ell} < r \le \frac{p-q}{\ell}\\
  3(1-\frac{\ell}{p}r), & \frac{p-q}{\ell} < r \le \frac{2q}{\ell}.
 \end{array}\right.
\end{equation}
For $\kappa \ge 3$, one can check that $3-\frac{\ell}{q}r \le 3\left(1-\frac{\ell}{p}r\right)$. Furthermore, if $3-\frac{\ell}{q}r < 2\left(1-\frac{\ell}{p}r\right)+\frac{q}{p}$, then $1-\frac{q}{p} < \frac{(p-2q)\ell}{pq}r$ which results in $r > \frac{(p-q)q}{(p-2q)\ell}$. One can see that
$\frac{(p-q)q}{(p-2q)\ell} \ge \frac{q}{\ell}$. Thus for
$\kappa \ge 3$
\begin{equation}
 d^*(r) = \left\{
 \begin{array}{ll}
  2(1-\frac{\ell}{p}r)+\frac{q}{p}, & \frac{q}{\ell} < r \le \frac{(p-q)q}{(p-2q)\ell}\\
  3-\frac{\ell}{q}r, &  \frac{(p-q)q}{(p-2q)\ell} < r \le \frac{2q}{\ell}.
 \end{array}\right.
\end{equation}

Now consider $\frac{2q}{p} < r \le \frac{p}{\ell}$. In this
region  $d(r)=
\min\{3(1-\frac{\ell}{p}r),2(1-\frac{\ell}{p}r)+\frac{q}{p},1+\frac{2q}{p}-\frac{\ell}{p}r\}$.
One can check that if $3\left(1-\frac{\ell}{p}r\right) <
2\left(1-\frac{\ell}{p}r\right)+\frac{q}{p}$, then $r
> \frac{p-q}{\ell}$.
Moreover, 
\begin{equation}\nonumber
 \text{if} ~~~3\left(1-\frac{\ell}{p}r\right) < 1-\frac{\ell}{p}r+\frac{2q}{p} \Rightarrow r > \frac{p-q}{\ell}.
\end{equation}
One can check that $\frac{p-q}{\ell} \le \frac{2q}{p}$ if and only if $\kappa \le 3$. Considering the fact that the cooperation is avoided whenever it is beneficial to do so, for $2 \le
\kappa \le 3$ we have
\begin{equation}
 d^*(r) = \left\{
 \begin{array}{ll}
   3(1-\frac{\ell}{p}r), & \frac{2q}{p} < r \le \frac{2p}{3\ell-p}\\
   1-r, & \frac{2p}{3\ell-p} \le r \le 1.
 \end{array}\right.
\end{equation}

For $\kappa \ge 3$ and for $\frac{2q}{\ell} < r \le \frac{p-q}{\ell}$, $3\left(1-\frac{\ell}{p}r\right) >
2\left(1-\frac{\ell}{p}r\right)+\frac{q}{p}$, and $3\left(1-\frac{\ell}{p}r\right) > 1+\frac{2q}{p}-\frac{\ell}{p}r$. In this region
\begin{equation}\nonumber
 \text{if}~~~ 2\left(1-\frac{\ell}{p}r\right)+\frac{q}{p} \le 1
 +\frac{2q}{p}-\frac{\ell}{p}r ~\Rightarrow~ r \ge \frac{p-q}{\ell}.
\end{equation}
Considering the fact that the cooperation is avoided whenever it is beneficial, for $\kappa \ge 3$ we have
\begin{equation}
 d^*(r) = \left\{
\begin{array}{cc}
 1+\frac{2q}{p}-\frac{\ell}{p}r, & \frac{2q}{\ell} \le r \le \frac{p-q}{\ell}\\
 3(1-\frac{\ell}{p}r), & \frac{p-q}{\ell} \le r \le \frac{2p}{3\ell-p}\\
 1-r, &  \frac{2p}{3\ell-p} \le r \le 1.
\end{array}\right.
\end{equation}
By summarizing the above results, the proof of the first part is concluded. For the proof of the second part, it is seen that $\kappa = 2$ provides the best diversity gain for
\begin{equation}\nonumber
 r \le \frac{q}{\ell} = \frac{q}{p+q} = \frac{1}{3}.
\end{equation}
The corresponding diversity gain in this region is given by
\begin{align}
 d^*(r) = 3\Big(1-\frac{\ell}{p}r\Big) = 3\Big(1-\frac{3}{2}r\Big), ~ 0 \le r \le \frac{1}{3}.
\end{align}

For other values of $r$ the maximum diversity gain is achieved
when $r = \frac{p-q}{\ell}$. In this case,
\begin{equation*}
 r = \frac{p-q}{\ell} = \frac{\kappa-1}{1+\kappa}.
\end{equation*}
Thus,
\begin{eqnarray}\nonumber
 \kappa = \frac{1+r}{1-r}.
\end{eqnarray}
For $\frac{1}{3} \le r \le \frac{1}{2}$, we obtain
\begin{align*}\nonumber
 d^*(r) = 2\left(1-\frac{\ell}{p}r\right)+\frac{q}{p} = \frac{3(1-r)}{1+r}.
\end{align*}
For $\frac{1}{2} \le r \le 1$, we obtain
\begin{equation*}\nonumber
 d^*(r) = 1-\frac{\ell}{p}r+\frac{2n}{p}
 = \frac{3(1-r)}{1+ r}.
\end{equation*}
By combining the results,we have
\begin{equation}
d^*(r) = \left\{
\begin{array}{cc}
  3\left(1-\frac{3}{2}r\right), & 0 \le r \le \frac{1}{3}\\
  \frac{3(1-r)}{1+r}, & \frac{1}{3} \le r \le 1.
\end{array}\right.
\end{equation}
The corresponding $\kappa$ is given by
\begin{equation}
 \kappa = \left\{
\begin{array}{cc}
 2, & 0 \le r \le \frac{1}{3} \\
 \frac{1+r}{1-r} , & \frac{1}{3} \le r \le 1.
\end{array}\right.
\end{equation}
This concludes the proof. \hfill$\blacksquare$

\section{Proof of Theorem \ref{Theorem_OSDF_MRelay}}\label{Proof_Theorem_OSDF_MRelay}
It is known that if $\kappa \le M$
\begin{equation}\nonumber
 d_{E_M}(r) = 1+M-\frac{\ell}{q}r, ~~~ 0 \le r \le \frac{p}{\ell},
\end{equation}
else if $\kappa \ge M$
\begin{equation}\nonumber
 d_{E_M}(r)=\left\{
\begin{array}{ll}
 1+M-\frac{\ell}{q}r, & 0 \le r \le \frac{Mq}{\ell}\\
 1+\frac{Mq}{p}-\frac{\ell}{p}r, & \frac{Mq}{\ell} \le r \le \frac{p}{\ell}.
\end{array}\right.
\end{equation}
In addition,
\begin{equation}\nonumber
 Pr(E_M) = \left\{
\begin{array}{ll}
 1, & 0 \le r \le \frac{p}{\ell}\\
 0, & \frac{p}{\ell} < r \le 1.
\end{array}\right.
\end{equation}
Let $b_m(r),~ m=0, \ldots, M$, be the negative exponent of $\rho$  in $Pr(E_m)\rho^{-d_{E_m}}$ when $\rho \rightarrow \infty$. The resulted DMT can be expressed as
\begin{equation}\nonumber
 d^*_M(r) = \min\left\{\left(1-\frac{\ell}{p}r\right)+d^*_{M-1}(r),b_M(r)\right\},
\end{equation}
which is simplified as follows.\\
If $\kappa \le M$
\begin{equation}\nonumber
d^*_M(r) = \min\Big\{\Big(1-\frac{\ell}{p}r\Big)+d^*_{M-1}(r),1+M-\frac{\ell}{q}r\Big\},~ 0 \le r \le \frac{p}{\ell},
\end{equation}
else, for $\kappa \ge M$,
\begin{align*}
d^*_M(r) =\left\{
\begin{array}{ll}
\min\Big\{(1-\frac{\ell}{p}r)+d^*_{M-1}(r),1+M-\frac{\ell}{q}r\Big\}, & 0 \le r \le \frac{Mq}{\ell}\\
\min\Big\{(1-\frac{\ell}{p}r)+d^*_{M-1}(r),1+\frac{Mq}{p}-\frac{\ell}{p}r\Big\},
 & \frac{Mq}{\ell} \le r \le \frac{p}{\ell}.
\end{array}\right.
\end{align*}

One can check
\begin{align*}
\text{If}~~ 1+M-\frac{\ell}{q}r \le b_0(r)
&\Rightarrow~ 1+M-\frac{\ell}{q}r \le (M+1)\Big(1-\frac{\ell}{p}r\Big)\\
 &\Rightarrow~ \kappa \ge M+1.
\end{align*}
In addition,
\begin{align*}
  \text{If}~~ 1+\frac{Mq}{p}-\frac{\ell}{p}r & \le  (M+1)\Big(1-\frac{\ell}{p}r\Big)\\
  & \Rightarrow r \le \frac{p-q}{\ell}
\end{align*}
Clearly, for $\kappa \le M+1$, $\frac{p-q}{\ell} \le \frac{Mq}{\ell}$.Hence, for $1 \le \kappa \le M+1$
\begin{equation}
 d_M^*(r) = \left(1-\frac{\ell}{p}r\right)+d^*_{M-1}(r), ~~~ 0 \le r \le \frac{p}{\ell}.
\end{equation}

For $\kappa \ge M+1$, when $0\le r \le \frac{(M-1)q}{\ell}$ we have
\begin{align*}
 b_{M-1}(r)&=\left(1-\frac{\ell}{p}r\right)+M-\frac{\ell}{q}r\\
  &\le 1+M-\frac{\ell}{q}r=b_M(r).
\end{align*}
Hence, for $\kappa \ge M+1$,
\begin{equation}\nonumber
  d_M^*(r) = \Big(1-\frac{\ell}{p}r\Big) + d_{M-1}^*(r), ~~~ 0 \le r \le \frac{(M-1)q}{\ell}.
\end{equation}

For $\frac{(M-1)q}{\ell} \le r \le
\frac{Mq}{\ell}$, we have
\begin{align}\nonumber
&b_0(r) = (M+1)\Big(1-\frac{\ell}{p}r\Big),\\ \nonumber
&b_1(r) = M\Big(1-\frac{\ell}{p}r\Big)+\frac{q}{\ell},\\ \nonumber
&b_2(r) = (M-1)\Big(1-\frac{\ell}{p}r\Big)+\frac{2q}{\ell},\\ \nonumber
& \hspace{3.5 cm} \vdots\\ \nonumber
&b_{M-1}(r) = 2\Big(1-\frac{\ell}{p}r\Big)+\frac{(M-1)q}{\ell},\\ \nonumber
&b_M(r) = 1+M-\frac{\ell}{q}r.
\end{align}
It can easily see that if $r \le \frac{p-q}{\ell}$, then $b_{M-1}(r) \le b_{M-2}(r) \le \ldots \le b_0(r)$ and vice versa. On the other hand, for $\kappa \ge M+1$, $\frac{p-q}{\ell} \ge \frac{Mq}{\ell}$. Hence, to determine the diversity gain when $\frac{(M-1)q}{\ell} \le r \le \frac{Mq}{\ell}$, only $b_M(r)$ and $b_{M-1}(r)$ need to be compared.
\begin{align}\nonumber
&\text{If}  ~~~ b_{M-1}(r) \le b_M(r)\\ \nonumber
&\Rightarrow~~2\Big(1-\frac{\ell}{p}r\Big)+\frac{(M-1)q}{\ell} \le 1+M-\frac{\ell}{q}r\\
&\Rightarrow~~ r \le \frac{(M-1)p^2q}{\ell^2(p-2q)}
\end{align}
Assuming $\eta_1 = \frac{(M-1)p^2q}{\ell^2(p-2q)}$, for $\kappa \ge M+1$, $\frac{(M-1)q}{\ell} \le \eta_1 \le \frac{Mq}{\ell}$.
Hence,
\begin{align}\nonumber
&d^*_M(r) = \\
&\left\{
\begin{array}{ll}
 2(1-\frac{\ell}{p}r)+\frac{(M-1)q}{p}, & \frac{(M-1)q}{\ell} \le r \le \eta_1\\
 1+M-\frac{\ell}{q}r, & \eta_1 \le r \le \frac{Mq}{\ell}.
\end{array}\right.
\end{align}
For $\frac{Mq}{\ell}\le r \le \frac{p}{\ell}$, we have
\begin{equation}
d^*_M(r) = \left\{
\begin{array}{cc}
 1+\frac{Mq}{\ell}-\frac{\ell}{p}r, & \frac{Mq}{\ell} \le r \le \frac{p-q}{\ell}\\
 (M+1)(1-\frac{\ell}{p}r), & \frac{p-q}{\ell} \le r \le \frac{p}{\ell}.
\end{array}\right.
\end{equation}
For $\frac{p}{\ell} \le r \le 1$, $d^*_M(r) = 0$. The resulted DMT in each region is compared to $(1-r)$ to
determine wether or not avoiding the cooperation. Proof of the second part of the Theorem is similar to the proof of the second part of Proposition \ref{Proposition_OSDF_2Relay}. \hfill$\blacksquare$

\section{Proof of Theorem \ref{Theorem_NAF_infinite_p}}\label{Apx_proof_Theorem_NAF_infinite_p}
The goal is to find $d^*(r)$ which is characterized by the following optimization problem.
\begin{equation*}
  d^*(r) = \inf_{(p-q)\alpha_0+q\min\{2\alpha_0-1,\beta\} > p-\ell r} \alpha_0+M\beta.
\end{equation*}
If $\min\{2\alpha_0-1,\beta\} = 2\alpha_0-1$, then $\beta \ge \max\{0,2\alpha_0-1\}$ and we get
\begin{align*}
  d^*(r) & = \inf_{\alpha_0 \ge 1-r} \alpha_0+M\max\{0, 2\alpha_0-1\}\\
  & = \left\{
  \begin{array}{ll}
    1-r & 0\le \alpha_0 \le \frac{1}{2}\\
    (1-r) + M(1-2r) & \frac{1}{2} \le \alpha_0 \le 1.
  \end{array}\right.\\
  & = \left\{
  \begin{array}{ll}
    (1-r) + M(1-2r) &  0\le r \le \frac{1}{2} \\
    1-r & \frac{1}{2}\le r \le 1.
  \end{array}\right.
\end{align*}

If $\min\{2\alpha_0-1,\beta\} = \beta$, then $1 \ge \alpha_0 \ge \frac{1+\beta}{2}$ and we get
\begin{align*}
  d^*(r) & = \inf_{(p-q)\alpha_0 + q\beta > p-\ell r} \alpha_0+M\beta
\end{align*}
If $r \ge p/\ell$, then $p-\ell r \le 0$. In this case, $\alpha_0=1/2, \beta = 0$ is the optimal solution. One can check that $\alpha_0=1/2, \beta = 0$ is also the optimal solution for $1/2 \le r \le p/\ell$. Hence,
\begin{equation*}
  d^*(r) = \frac{1}{2}, ~~~~ \frac{1}{2} \le r \le 1.
\end{equation*}

For $0 \le r \le 1/2$, if $p=q$, we obtain
\begin{align*}
  d^*(r) & = \inf_{\beta \ge \max\{0,1-2r\}} \frac{1+\beta}{2}+M\beta\\
  & = 
    (1-r) + M(1-2r) & 0 \le r \le \frac{1}{2}.
\end{align*}

For $0 \le r \le 1/2$ and $p \ne q$, the cross point of the two linear conditions, $\alpha_0 = 1-r, \beta = 1-2r$, is a feasible solution. The objective value for this solution is
\begin{equation*}
  d(r) = (1-r) + M(1-2r), ~~~ 0 \le r \le \frac{1}{2}.
\end{equation*}
One can check that $\alpha'_0 = 1-r+\delta$ and $\beta' = 1-2r-\frac{p-q}{q}\delta$, for a positive value of $\delta$, is also a feasible solution if
\begin{align*}
  \delta & \le \min\left\{r, \frac{q}{p-q}(1-2r)\right\}\\
  & = \left\{
  \begin{array}{ll}
    r, & 0 \le r \le \frac{q}{\ell}\\
    \frac{q}{p-q}(1-2r), & \frac{q}{\ell} \le r \le \frac{1}{2}.
  \end{array}\right.
\end{align*}
The above condition comes from the fact that $\alpha'_0 \le 1$ and $\beta' \ge 0$. The objective value for the new feasible solution is
\begin{equation*}
  d(r) = (1-r) + M(1-2r) + \delta\Big(1-\frac{M(p-q)}{q}\Big), ~~~ 0 \le r \le \frac{1}{2}
\end{equation*}
It is seen that for $\kappa \le \frac{M+1}{M}$, the term $\big(1-\frac{M(p-q)}{q}\big)$ is positive and it increases the objective value for any positive value of $\delta$. Hence, $\alpha_0 = 1-r, \beta = 1-2r$ is in fact the optimum solution for $1 < \kappa \le \frac{M+1}{M}$ and we get
\begin{equation*}
d^*(r) = \left\{
\begin{array}{ll}
  (1-r) + M(1-2r), & 0 \le r \le \frac{1}{2}\\
  \frac{1}{2}, & \frac{1}{2} \le r \le 1.
\end{array}\right.
\end{equation*}

For $\kappa \ge \frac{M+1}{M}$, the term $\big(1-\frac{M(p-q)}{q}\big)$ is negative and it decreases the objective value for any positive value of $\delta$. The optimal solution which is achieved for the maximum value of $\delta$ in each region is given by
\begin{equation*}
  d^*(r) = \left\{
  \begin{array}{ll}
    1-\frac{M(p-q)}{q}r + M(1-2r), & 0 \le r \le \frac{q}{\ell}\\
    (1-r)+\frac{q}{p-q}(1-2r), & \frac{q}{\ell} \le r \le \frac{1}{2}\\
    \frac{1}{2}, & \frac{1}{2} \le r \le 1.
  \end{array}\right.
\end{equation*}

By comparing the results for different values of $\kappa$, it is seen that the best DMT is obtained when $1 \le \kappa \le \frac{M+1}{M}$ and is given by
\begin{equation*}
  d^*(r) = (1-r) + M(1-2r)^+, ~~~ 0 \le r \le 1.
\end{equation*}
This concludes the proof. \hfill$\blacksquare$

\section{proof of Theorem \ref{Theorem_MainResult_M_Relay_NAF}}\label{Apx_Proof_Theorem_NAF_finite_u}
The goal is to solve the following optimization problem.
\begin{equation*}
  d^*(r) = \inf_{p\alpha_0 + q \min\{\alpha_0,\beta\} > \ell(1-r)} \alpha_0+M\beta.
\end{equation*}

If $\min\{\alpha_0,\beta\} = \alpha_0$, then
\begin{align}\nonumber
  d^*(r) &= \inf_{\alpha_0 > 1-r} (M+1)\alpha_0\\ \nonumber
  &= (M+1)\left(1-r\right).
\end{align}

If $\min\{\alpha_0,\beta\} = \beta$, in this case $\hat{\alpha}_0 = \hat{\beta} = 1-r$ is a feasible solution. The objective value for this feasible solution is $d(r) = (M+1)(1-r)$.

Let $\tilde{\alpha}_0 \triangleq \hat{\alpha}_0 + \delta$, where $\delta$ is a positive real number. In this case $\tilde{\alpha}_0$ and $\tilde{\beta} = \hat{\beta} - \frac{p}{q}\delta$ is another feasible solution. The objective value for the new variables is
\begin{equation}\label{dr-variable}
  d^*(r) = (M+1)\left(1-r\right)-\left(\frac{Mp}{q}-1\right)\delta.
\end{equation}
As $0 \le \beta \le \alpha_0 \le 1$, $\delta$ should be chosen such that $\tilde{\alpha}_0 \le 1$, and $\tilde{\beta} \ge 0$. We get
\begin{eqnarray}\nonumber
  \tilde{\alpha}_0 \le 1 &\rightarrow& \delta \le r\\ \nonumber
  \tilde{\beta} \ge 0 &\rightarrow& \delta \le \frac{q}{p}\left(1-r\right).
\end{eqnarray}
As both conditions should be satisfied, we have
\begin{equation}\nonumber
  \delta = \left\{
  \begin{array}{cc}
    r, & 0 \le r \le \frac{q}{\ell}\\ \nonumber
    \frac{q}{p}(1-r) , & \frac{q}{\ell} \le r \le 1.
  \end{array}\right.
\end{equation}
By replacing $\delta$ into (\ref{dr-variable}), we obtain
\begin{equation}\nonumber
  d^*(r) = \left\{
  \begin{array}{cc}
    (M+1)(1-\frac{M\ell}{(M+1)n}r), & 0 \le r \le \frac{q}{\ell}\\
    1+\frac{q}{p}-\frac{\ell}{p}r, & \frac{q}{\ell} \le r \le 1.
  \end{array}\right.
\end{equation}
One can see that the best DMT is achieved when $\kappa = 1$. This concludes the proof. \hfill $\blacksquare$

\bibliographystyle{ieeetr} 
\bibliography{Asynchronous_DFAF_JOurnalVersion}
\end{document}